\documentclass[11pt]{article}
\usepackage[T1]{fontenc}
\usepackage{mathtools}
\usepackage{amssymb}
\usepackage{stmaryrd}
\usepackage[mathscr]{euscript}
\usepackage{dsfont}
\usepackage{slashed}
\usepackage{fullpage}
\usepackage{currvita}
\usepackage{tocloft}
\usepackage{comment}
\usepackage[numbers,sort,compress]{natbib}
\usepackage{array}
\newcolumntype{L}{>{$}l<{$}}
\usepackage{pict2e}

\usepackage[colorlinks=true,linkcolor=blue,citecolor=magenta,linktocpage=true]{hyperref}
\usepackage{braket}
\usepackage{titlesec}
\titleformat*{\section}{\normalsize\bfseries}
\titleformat*{\subsection}{\normalsize\bfseries}
\titleformat*{\subsubsection}{\normalsize\bfseries}
\makeatletter
\renewcommand{\@dotsep}{1000}
\def\be#1\ee{\begin{align}#1\end{align}}
\def\bsub#1\esub{\begin{subequations}#1\end{subequations}}

\def\q{\qquad}
\def\f{\frac}
\def\eps{\varepsilon}

\def\ip{\lrcorner\,}
\def\ipp{\hbox{$\ip\!\!\!\!\!\;\ip$}}

\def\de_\omega{\mathrm{D}}
\def\de{\mathrm{d}}

\def\B{\mathcal{B}}
\def\C{\mathcal{C}}

\def\I{\mathcal{I}}
\def\J{\mathcal{J}}

\def\O{\mathcal{O}}

\def\Q{\mathcal{Q}}

\def\T{\mathcal{T}}
\def\U{\mathcal{U}}
\def\V{\mathcal{V}}
\def\W{\mathcal{W}}

\numberwithin{equation}{section}

\DeclareRobustCommand{\loplus}{\mathbin{\mathpalette\dog@lsemi{+}}}
\DeclareRobustCommand{\lotimes}{\mathbin{\mathpalette\dog@lsemi{\times}}}
\DeclareRobustCommand{\roplus}{\mathbin{\mathpalette\dog@rsemi{+}}}
\DeclareRobustCommand{\rotimes}{\mathbin{\mathpalette\dog@rsemi{\times}}}

\newcommand{\dog@rsemi}[2]{\dog@semi{#1}{#2}{-90,90}}
\newcommand{\dog@lsemi}[2]{\dog@semi{#1}{#2}{270,90}}
\newcommand{\dog@semi}[3]{%
  \begingroup
  \sbox\z@{$\m@th#1#2$}%
  \setlength{\unitlength}{\dimexpr\ht\z@+\dp\z@\relax}%
  \makebox[\wd\z@]{\raisebox{-\dp\z@}{%
    \begin{picture}(1,1)
    \linethickness{\variable@rule{#1}}
    \roundcap
    \put(0.5,0.5){\makebox(0,0){\raisebox{\dp\z@}{$\m@th#1#2$}}}
    \put(0.5,0.5){\arc[#3]{0.5}}
    \end{picture}%
  }}%
  \endgroup
}
\newcommand{\variable@rule}[1]{%
  \fontdimen8  
  \ifx#1\displaystyle\textfont3\else
    \ifx#1\textstyle\textfont3\else
      \ifx#1\scriptstyle\scriptfont3\else
        \scriptscriptfont3\relax
  \fi\fi\fi
}
\makeatother

\begin{document}

\title{\Large{\textbf{\sffamily 3d gravity in Bondi--Weyl gauge:\\
charges, corners, and integrability}}}
\author{\sffamily Marc Geiller$^1$, Christophe Goeller$^2$, C\'eline Zwikel$^3$}
\date{}
\date{\small{\textit{$^1$Univ Lyon, ENS de Lyon, Univ Claude Bernard Lyon 1,\\ CNRS, Laboratoire de Physique, UMR 5672, F-69342 Lyon, France\\ $^2$Arnold Sommerfeld Center for Theoretical Physics,\\ Ludwig-Maximilians-Universit\"at M\"unchen,\\
Theresienstrasse 37, 80333 M\"unchen, Germany\\ $^3$Institute for Theoretical Physics, TU Wien,\\ Wiedner Hauptstrasse 8–10/136, A-1040 Vienna, Austria\\}}}

\maketitle

\begin{abstract}
We introduce a new gauge and solution space for three-dimensional gravity. As its name Bondi--Weyl suggests, it leads to non-trivial Weyl charges, and uses Bondi-like coordinates to allow for an arbitrary cosmological constant and therefore spacetimes which are asymptotically locally (A)dS or flat.
%Turning on the Weyl charges requires leaky boundary conditions and an unspecified variational principle.
We explain how integrability requires a choice of integrable slicing and also the introduction of a corner term. After discussing the holographic renormalization of the action and of the symplectic potential, we show that the charges are finite, symplectic and integrable, yet not conserved. We find four towers of charges forming an algebroid given by $\mathfrak{vir}\oplus\mathfrak{vir}\oplus\text{Heisenberg}$ with three central extensions, where the base space is parametrized by the retarded time. These four charges generate diffeomorphisms of the boundary cylinder, Weyl rescalings of the boundary metric, and radial translations.
%It is the first time that four towers of finite charges appear in three-dimensional gravity.
We perform this study both in metric and triad variables, and use the triad to explain the covariant origin of the corner terms needed for renormalization and integrability.
\end{abstract}

\thispagestyle{empty}
\newpage
\setcounter{page}{1}

\tableofcontents
\bigskip
\hrule

\section{Motivations}

Suitable gravitational asymptotic symmetries are diffeomorphisms which preserve a certain set of conditions on the metric (typically boundary or fall-off conditions together with a choice of bulk gauge) and furthermore lead to finite and non-trivial surface charges.  Their study is of particular importance because they endow bounded spacetime regions with charges and a charge algebra, which contain important physical information about the classical theory (such as the observables and potential radiative degrees of freedom), and even about the IR regime of the quantum theory \cite{Strominger:2013jfa,He:2014laa,Strominger:2017zoo,Ashtekar:2018lor}. 
%This has motivated a significant body of work on asymptotic symmetries in four-dimensional gravity \cite{Bondi:1962px,Sachs:1962wk,Sachs:1962zza,Barnich:2010eb,Barnich:2011mi,Compere:2020lrt,Fiorucci:2020xto,Ruzziconi:2020cjt,Freidel:2021yqe}, and also on its lower-dimensional relatives.

It has long been recognized that three-dimensional gravity, in spite of being topological, is a non-trivial setup where asymptotic symmetries also play an important role. Since the seminal work of Brown--Henneaux on AdS$_3$ spacetimes with eponymous boundary conditions \cite{Brown:1986nw}, which has revealed a double Virasoro algebra of asymptotic charges and sparked the development of AdS/CFT, numerous alternative gauge choices, boundary conditions and locations of the said boundary have been studied, see for example \cite{Ashtekar:1996cd,Barnich:2006av,Compere:2008us,Barnich:2012aw,Troessaert:2013fma,Compere:2014cna,Donnay:2015abr,Compere:2015knw,Donnay:2016ejv,Afshar:2016wfy,Perez:2016vqo,Grumiller:2016pqb,Oblak:2016eij,Grumiller:2017sjh,Grumiller:2019fmp,ruzziconi2020conservation,Alessio:2020ioh,Adami:2020ugu}. Aside from the richness of three-dimensional gravity, this abundance of works reveals that there is a large freedom (sometimes even referred to as an art) in choosing the gauge and boundary conditions, and that the big picture connecting all these choices is far from being understood.

The question we address in this work is (in part) that of understanding the Weyl rescalings of the boundary metric. In Penrose's geometrical treatment of null infinity via the conformal compactification, the boundary geometry is characterized by an equivalence class of metrics under local conformal rescalings \cite{Penrose:1962ij,Penrose:1964ge}. In coordinate descriptions of asymptotic boundaries, following e.g. the Bondi--Sachs formalism \cite{Bondi:1962px,Sachs:1962zza,Sachs:1962wk} (or the Fefferman--Graham gauge for the AdS$_3$ boundary \cite{2007arXiv0710.0919F}), the boundary metric is however usually frozen and no Weyl rescaling is allowed. This is for example what happens with the Brown--Henneaux Dirichlet boundary conditions leading to the double Virasoro algebra \cite{Brown:1986nw}, or with the Bondi gauge metrics leading to typical realizations of the $\mathfrak{bms}_3$ algebra \cite{Barnich:2006av,Barnich:2012aw}.

While the role of the Weyl transformations has been discussed extensively in the literature, in particular in relation to holography \cite{Henningson:1998gx,Bautier:1999ic,Imbimbo:1999bj,Schwimmer:2000cu,Rooman:2000zi,Kalkkinen:2001vg,Papadimitriou:2005ii,Troessaert:2013fma,Ciambelli:2019bzz,Alessio:2020ioh,Ciambelli:2020ftk,Ciambelli:2020eba}, only a few works have been devoted to the detailed study of the Weyl charges. The few available studies of the Weyl charges have been done in different spacetime dimensions and with different choices of gauge. We summarize briefly these results in three and four dimensions to contrast them with those of the present paper (which we will explain below).
\begin{itemize}
\item In \cite{Compere:2020lrt} the authors have used the Starobinsky/Fefferman--Graham gauge in four dimensions with a free boundary metric (i.e. without imposing a variational principle), and also constructed the flat limit using a diffeomorphism leading to the Bondi gauge \cite{Poole:2018koa,Compere:2019bua,Ruzziconi:2020cjt}. They have found vanishing Weyl charges in the Starobinsky/Fefferman--Graham gauge. In \cite{Fiorucci:2020xto} the authors have extended the analysis of \cite{Compere:2020lrt} to arbitrary spacetime dimensions, and found that the Weyl charges are non-vanishing in the odd-dimensional case. In order to obtain non-vanishing Weyl charges in the four-dimensional case, \cite{Barnich:2010eb,Barnich:2016lyg,Barnich:2019vzx} have proposed a relaxed condition on the metric of the celestial 2-sphere, and studied the asymptotic symmetries in this new gauge (without however computing the associated charges). The authors of \cite{Freidel:2021yqe} have built upon this proposal by computing the renormalized charges, and proposed a generalized BMS--Weyl algebra. Weyl charges were also found in \cite{Adami:2020amw,progress-1} when studying null boundaries at finite distance.
\item In \cite{Troessaert:2013fma,Alessio:2020ioh} the authors have used the Fefferman--Graham gauge in three dimensions, and relaxed just enough conditions on the boundary metric to allow for non-vanishing Weyl charges. In \cite{ruzziconi2020conservation} the authors have used the Bondi gauge in three dimensions with a free boundary metric, and found that the Weyl charges were vanishing. This result is surprising because, as mentioned above, non-trivial Weyl charges have been found in \cite{Alessio:2020ioh,Fiorucci:2020xto} using the Fefferman--Graham gauge, and also in \cite{Adami:2020ugu} when studying null boundaries at finite distance.
\end{itemize}

In the present work we generalize the results of \cite{ruzziconi2020conservation} and propose a new gauge, called Bondi--Weyl gauge, which allows to construct non-trivial Weyl charges in three-dimensional spacetimes. As the name also suggests, our analysis is performed in Bondi-like coordinates so as to allow for an arbitrary cosmological constant and to have a well-defined flat limit. In this setup the boundary is at $r=\infty$. An important and subtle point for the study of the Weyl charges is to allow for so-called leaky boundary conditions, which is realized by unfreezing the boundary metric. As we leave the boundary dynamics unspecified, this means in particular that the variational principle is also unspecified. In addition to this, we relax the so-called determinant condition by allowing a subleading term in the celestial metric. Importantly, this relaxation is stronger than the one suggested in \cite{Barnich:2010eb,Freidel:2021cbc}. It is precisely this new subleading term which is responsible for the appearance of the Weyl charges.

The use of leaky boundary conditions implies that the symplectic potential and the charges are generically divergent at $\I^+$. This requires a procedure of symplectic renormalization, which here follows closely that used in \cite{ruzziconi2020conservation}. In addition, the use of leaky boundary conditions implies that the charges are a priori non-integrable and non-conserved. The former property of leaky boundary conditions has been related in \cite{Adami:2020ugu} to the presence of propagating degrees of freedom passing through the boundary. However, since we are here in three-dimensional gravity, this non-integrability is not due to physical symplectic flux, and can be bypassed with a so-called change of slicing, which corresponds to a field-dependent redefinition of the vector field generating the asymptotic symmetries. As a novelty, we also find that integrability requires to add a so-called corner ambiguity to the symplectic potential.\footnote{This is similar to the case studied in \cite{Adami:2021sko} where generic boundary conditions around a finite null surface were worked out in Topologically Massive Gravity. Therein, it was shown that it exists specific choices of slicing and of a corner ambiguity for which the non-integrability of the charges is only sourced by the physical flux passing through the boundary.}

At the end of the day, after symplectic renormalization, change of slicing, and the introduction of the proper corner term, we find the integrable charges \eqref{final charge}. Moreover these charges are symplectic, i.e. independent of $r$ and therefore actually defined at any finite distance. Because of the leaky boundary conditions these charges are non-conserved in spite of being integrable. The surprising result is that these charges have four components (or towers), forming in the integrable slicing the centrally-extended algebra $\mathfrak{vir}\oplus\mathfrak{vir}\oplus\text{Heisenberg}$. The component $\mathfrak{vir}\oplus\mathfrak{vir}$ is generated by the two $(u,\phi)$-dependent conformal generators (or the $(u,\phi)$-dependent superrotations and supertranslations in the flat limit), while the Heisenberg component is generated by the conformal factor of the boundary metric and the subleading term introduced in our new determinant condition. Geometrically, these Heisenberg generators describe Weyl transformations of the boundary metric and radial translations.

With this new Bondi--Weyl gauge, we therefore obtain four unconstrained time-dependent finite and integrable charges. Until now, the maximal number of unconstrained boundary degrees of freedom which had been exhibited was three \cite{Grumiller:2020vvv,Adami:2020ugu} (we note however that six constrained degrees of freedom were obtained in the Chern--Simons formulation in \cite{Grumiller:2016pqb,Grumiller:2017sjh}). The difference with previous work, which allows for the appearance of these four charges, is two-fold. First, the boundary metric is freely fluctuating, and consequently the charges have an unconstrained time evolution reflected in an arbitrary $u$-dependency. Second, the metric is allowed to have divergent and subleading components, which however at the end of the day end up contributing in a finite manner to the charges. These two ingredients could potentially be used to generalize other proposals for boundary conditions and thereby obtain new boundary charges.

In addition to the introduction of the new Bondi--Weyl gauge and the study of the renormalization and integrability of the charges, which we initially perform in the metric formulation, we explain in details how the analysis can be performed in the triad formulation. This requires a careful analysis of the symmetries acting on the triad, which are diffeomorphisms improved by Lorentz transformations, and of the corner term relating the metric and triad formulations \cite{DePaoli:2018erh,Oliveri:2019gvm,Oliveri:2020xls,Freidel:2020xyx,Freidel:2020svx}. Additional subtleties arise here because of the corner terms needed for symplectic renormalization and integrability. This is of particular interest because the study of the triad formulation actually allows to understand the origin of the corner terms needed for renormalization and integrability. 

\subsection*{Organization of the paper}
We start in section \ref{sec:gauge} by defining the Bondi--Weyl gauge. We detail its relation with the usual Bondi gauge, work out the solution space, and compute its gauge-preserving residual symmetries. Next, in section \ref{sec:renormalization} we tackle the computation of the charges associated with these residual symmetries. This requires to consider corner ambiguities in the definition of the symplectic potential in order to obtain finite and integrable charges. More precisely, we first discuss the renormalization of the action, and then that of the symplectic potential. We also comment on the relationship between the Lagrangian boundary term and the corner ambiguities in the potential (and elaborate more on this in section \ref{sec:origin}). Then, we exhibit the corner terms needed for renormalization and integrability. We explain how integrability requires in addition a change of slicing (i.e. a field-dependent redefinition of the residual symmetries). We finally compute the charge algebra, and end the section by studying the limiting case of Dirichlet boundary conditions. In section \ref{sec:triad} we explain how to compute the charges in the triad formulation. This requires to study the symmetries of the triad and to introduce a relative corner term relating the metric and triad formulations. We then explain in section \ref{sec:origin} how this relative corner term between the triad and metric formulations actually gives rise to the renormalization and integrability corner terms which we have used to compute the charges. The twist is that this relative corner term has to be used in a counter-intuitive manner: it must be added in \textit{both} the metric and triad formulations in order to obtain finite and integrable charges. We finally conclude and give perspectives for future work in section \ref{sec:conclusion}.

\section{Bondi--Weyl gauge}\label{sec:gauge}

In this first section we introduce the Bondi--Weyl gauge, and then solve the Einstein field equations in this gauge to find the on-shell metrics constituting our solution space. Once this is established, we find the asymptotic Killing vectors generating the residual symmetries preserving the solution space.

\subsection{Gauge choice}

We start in accordance with the usual Bondi--Sachs formalism \cite{Bondi:1962px,Sachs:1962zza,Sachs:1962wk}, and choose coordinates $(u,r,\phi)$ and metrics such that $\partial_\mu u$ is null and $\phi$, to be used as an angular coordinate, is constant along null rays. This means that $g^{\mu\nu}(\partial_\mu u)(\partial_\nu u)=0=g^{\mu\nu}(\partial_\mu u)(\partial_\nu\phi)$, which implies $g^{uu}=0=g^{u\phi}$, and in turn $g_{rr}=0=g_{r\phi}$. Using these two conditions puts the line element in the form
\be\label{metric}
\de s^2=\f{\V}{r}\B\,\de u^2-2\B\,\de u\,\de r+r^2\W(\de\phi-\U\de u)^2,
\ee
where the four functions $\B(u,r,\phi)$, $\U(u,r,\phi)$, $\V(u,r,\phi)$ and $\W(u,r,\phi)$ a priori depend on all three coordinates. At this point we have only used two conditions to put the metric in the form \eqref{metric}. This is where our gauge differs from previous proposals in the literature, as we now explain.

In the original Bondi--Sachs formalism \cite{Bondi:1962px,Sachs:1962zza,Sachs:1962wk}, one imposes (in $d$ spacetime dimensions with $d-2$ angular coordinates $A,B,\dots$) the so-called determinant condition $\det(g_{AB})=r^{2(d-2)}\det(g^\circ_{AB})$, where $g^\circ_{AB}$ is the unit sphere metric. In order to allow for Weyl rescalings of the transverse boundary metric, starting with \cite{Barnich:2010eb} several authors have proposed to work with the relaxed condition
\be\label{relaxed det}
\partial_r\left(\f{\det(g_{AB})}{r^{2(d-2)}}\right)=0,
\ee
which is equivalent to $\det(g_{AB})=r^{2(d-2)}e^{2\varphi}$ for some arbitrary function $\varphi(u,x^A)$. When $d=3$, this general Bondi gauge was studied in \cite{Ciambelli:2020ftk,Ciambelli:2020eba,ruzziconi2020conservation}. However, although it allows to describe Weyl rescalings of the transverse boundary metric, it leads to vanishing Weyl charges \cite{ruzziconi2020conservation}. This is in part our motivation to introduce an even more general gauge, called Bondi--Weyl gauge, in which we now consider a relaxed determinant condition by allowing
\be\label{new relaxed det}
\partial_r\left(\f{g_{\phi\phi}}{r^2}\right)=\partial_r\W
\ee
to be non-vanishing. As we are about to see, the Einstein equations will indeed fix a non-trivial $r$-dependency in the function $\W$. At the off-shell level however, since we are not imposing any condition on the angular part of the metric, we need to pick a third gauge-fixing condition. This will be the restriction $\B=e^{2\beta(u,\phi)}$, which achieves to determine our off-shell line element. Notice that here we are allowing for an arbitrary angular metric (at least off-shell), and imposing a condition on $\B$ by hand. This is to be contrasted with the choice made in \cite{Barnich:2010eb,Barnich:2012aw,Ciambelli:2020ftk,Ciambelli:2020eba,ruzziconi2020conservation} (and related work using either the original Bondi--Sachs determinant condition or the relaxed condition \eqref{relaxed det}), where $\partial_r\W=0$ is imposed off-shell but $\partial_r\B=0$ is only derived as an on-shell condition.

\subsection{Solution space}

Having introduced the off-shell line element \eqref{metric}, we can now go ahead and study the Einstein field equations $E_{\mu\nu}\equiv R_{\mu\nu}+2\ell^{-2}g_{\mu\nu}=0$. We have already fixed the radial dependency of $\B$ by hand with our choice of gauge. It turns out that now three of the six Einstein equations fix the radial dependency of $\W$, $\U$ and $\V$ in the following order:
\bsub
\be
E_{rr}=0\ &\Rightarrow\ \W=e^{2\varphi}\left(1-\f{H}{r}\right)^2,\label{W solution}\\
E_{r\phi}=0\ &\Rightarrow\ \U=U+\f{e^{2(\beta-\varphi)}}{(r-H)^2}\big(2(r-2H)\beta'-N\big),\\
E_{\phi\phi}=0\ &\Rightarrow\ \f{\V}{r}=2M-2r(U'+U\varphi'+\partial_u\varphi)-\f{e^{2\beta}}{\ell^2}r(r-2H)\cr
&\phantom{\ \Rightarrow\ \f{\V}{r}=}+\f{e^{2(\beta-\varphi)}}{(r-H)^2}(N+2H\beta')\big(2(2r-3H)\beta'-N\big).
\ee
\esub
Here prime is the derivative along $\phi$ and $\beta(u,\phi)$, $\varphi(u,\phi)$, $U(u,\phi)$, $H(u,\phi)$, $M(u,\phi)$ and $N(u,\phi)$ are arbitrary functions. We can think of these as ``boundary data'' since they only depend on the two coordinates $u$ and $\phi$, and will later on parametrize the charges. With this radial expansion, the component $E_{ur}$ of the field equations is then automatically satisfied.
 
The last two Einstein equations determine the evolution of $M$ and $N$, which are the Bondi mass and angular momentum aspect. More precisely, the component $E_{u\phi}$ is satisfied once we impose the additional constraint $\partial_uN=F_N$, while finally $E_{uu}$ is then solved once we impose $\partial_uM=F_M$, where $F_N$ and $F_M$ are lengthy functions of the boundary data $(\beta,\varphi,U,H,M,N)$ and their derivatives. On-shell, our solution space is therefore described by the functions $(\beta,\varphi,U,H,M,N)$ and two evolution constraints on $(M,N)$.

We have explained above that off-shell our Bondi--Weyl gauge is different from the Bondi gauge of \cite{Barnich:2010eb,Barnich:2012aw,Ciambelli:2020ftk,Ciambelli:2020eba,ruzziconi2020conservation}, because in this latter $\B$ is free while $\W$ is determined, whereas here we have made the opposite choice. However, on-shell, we simply recover an extension of the Bondi gauge by the function $H$. On-shell, we now call Bondi the metric with $H=0$ and Bondi--Weyl the metric with $H\neq0$ which we set out to study.

Now that we have solved the Einstein equations and obtained the on-shell metrics, we can discuss some aspects of their geometry. First, one can see that the relaxed determinant condition is now
\be
\partial_r\left(\f{g_{\phi\phi}}{r^2}\right)=\partial_r\W=\f{2}{r^3}e^{2\varphi}H(r-H),
\ee
and therefore controlled by the new function $H$ in our solution space. Since $H$ enters $\W$ in \eqref{W solution} as a subleading term, we see here that the determinant condition is satisfied asymptotically. Therefore, our Bondi--Weyl gauge is still consistent with the conformal compactification. The boundary metric, on the other hand, is $H$-independent and given by
\be\label{boundary metric}
\de\bar{s}^2=\bar{g}_{\mu\nu}\de x^\mu\,\de x^\nu\coloneqq\lim_{r\to\infty}\left(\f{\de s^2}{r^2}\right)=-\f{e^{4\beta}}{\ell^2}\de u^2+e^{2\varphi}(\de\phi-U\de u)^2.
\ee
This boundary metric is therefore allowed to freely fluctuate if $(\beta,\varphi,U)$ are arbitrary. In this sense the function $H$ parametrizes subleading corrections only.
%The metric \eqref{metric} has two natural null vectors, given by
%\be
%n^\mu=-\partial_\mu u,
%\q\q
%v^\mu=\B\left(\f{1}{2r}\V\partial_\mu u-\partial_\mu r\right),
%\q\q
%g_{\mu\nu}n^\mu v^\nu=-1.
%\ee
%Defining $q_{\mu\nu}=g_{\mu\nu}+n_\mu v_\nu+v_\mu n_\nu$, their twist is $H$ dependent and given by
%In the flat limit $\ell\to\infty$, the vector $v$ is generating the null surface at $r\to\infty$, and with these limits its surface gravity is
%\be
%\kappa=(\partial_u+U\partial_\phi)(2\beta-\varphi)-U'.
%\ee

Let us end this part with an interesting observation concerning the role of the new function $H$. If we perform a finite diffeomorphism $r\mapsto\bar{r}-H$ on the on-shell metric \eqref{metric}, we obtain the metric in Bondi gauge (i.e. with $H=0$) where the mass and angular momentum aspect are
\bsub\label{large diffeo}
\be
\bar{M}&=M-(UH)'-\partial_uH-H(U\varphi'+\partial_u\varphi)+\f{e^{2\beta}}{2\ell^2}H^2,\label{M bar}\\
\bar{N}&=N+2H\beta'-H',
\ee
\esub
and have therefore absorbed the dependency on $H$. This is the finite diffeomorphism which, starting from the Bondi gauge, shifts the radial coordinate $r$ by an arbitrary $(u,\phi)$-dependent function $H$ to reach the Bondi--Weyl gauge. As we are going to see, this diffeomorphism is large and $H$ leads to an independent charge.

\subsection{Residual symmetries}

We now search for the asymptotic Killing vectors $\xi^\mu=(\xi^u,\xi^r,\xi^\phi)$ which preserve our family of on-shell metrics. These can be found by solving successively components of the Killing equation as follows:
\bsub
\be
(\pounds_\xi g)_{rr}=0\ &\Rightarrow\ \xi^u=f,\\
(\pounds_\xi g)_{r\phi}= 0\ &\Rightarrow\ \xi^\phi=g-\f{e^{2(\beta-\varphi)}}{r-H}f',\\
\partial_r\big((\pounds_\xi g)_{ru}\big)=0\ &\Rightarrow\ \xi^r=\xi^r_0+r\xi^r_1-e^{2(\beta-\varphi)}\f{N+2H\beta'}{r-H}f',
\ee
\esub
where $f(u,\phi)$, $g(u,\phi)$, $\xi^r_0(u,\phi)$ and $\xi^r_1(u,\phi)$ are four free integration functions. For later convenience, we now change the free functions and introduce $h(u,\phi)$ by redefining
\be
\xi^r_1=h+Uf'-g'-g\varphi'-f\partial_u\varphi.
\ee
This has the advantage of giving to $\varphi$ the canonical transformation law $\delta_\xi\varphi=h$ as we are about to see. Finally, by looking at the Bondi limit where $H=0$, one can see that there is an unwanted $\O(r)$ piece in $(\pounds_\xi g)_{\phi\phi}\big|_{H=0}$, which can be removed by redefining the function 
\be
\xi^r_0=e^{2(\beta-\varphi)}\left(f''+f'(4\beta-\varphi)'\right)-k-H\xi^r_1,
\ee
where the new free function $k(u,\phi)$ is such that $k\big|_{H=0}=0$, and the shift by $H$ is simply for later convenience. We have now determined a parametrization of the asymptotic Killing vectors of the metric \eqref{metric}, which are given by
\bsub\label{AKV}
\be
\xi^u&=f,\\
\xi^\phi&=g-\f{e^{2(\beta-\varphi)}}{r-H}f',\\ \nonumber
\xi^r&=(r-H)(h+Uf'-g'-g\varphi'-f\partial_u\varphi)-k\\
&\phantom{=\ }+\left(f''+f'(4\beta-\varphi)'-\f{N+2H\beta'}{r-H}f'\right),
\ee
\esub
where $f(u,\phi)$, $g(u,\phi)$, $h(u,\phi)$ and $k(u,\phi)$ are arbitrary functions which, importantly for the change of slicing studied in section \ref{sec:slicing}, can be field-dependent. These asymptotic Killing vectors coincide with the ones derived in Bondi gauge \cite{Ciambelli:2020ftk,Ciambelli:2020eba} when $H=0$.

By construction, these vector fields preserve our on-shell family of metrics in the sense that $\pounds_\xi\big(g_{\mu\nu}(\Phi)\big)=g_{\mu\nu}(\delta_\xi\Phi)$, where $\Phi=\{\beta,\varphi,U,H,M,N\}$ denotes the set of fields. Explicitly, these fields transform as
\bsub\label{variations}
\be
\delta_\xi\beta&=f\partial_u\beta+g\beta'+Uf'+\f{1}{2}(\partial_uf-f\partial_u\varphi-g'-g\varphi'+h),\\
\delta_\xi\varphi&=h,\\
\delta_\xi U&=f\partial_uU+gU'+U(\partial_uf-g'+Uf')-\partial_ug+\f{e^{2(2\beta-\varphi)}}{\ell^2}f',\\
\delta_\xi H&=f\partial_uH+gH'+k,
\ee
\esub
and the variations $\delta_\xi M$ and $\delta_\xi N$ are too lengthy to be reproduced here, but are given in appendix \ref{app:variations}. With our Bondi--Weyl gauge and the relaxed determinant condition we find that
\be
\f{1}{2}g^{\phi\phi}(\pounds_\xi g)_{\phi\phi}=\delta_\xi\varphi-\f{\delta_\xi H}{r-H},
\ee
which again obviously generalizes the results in Bondi gauge.

Using the modified Lie bracket \cite{Barnich:2010eb},\footnote{This bracket is sometimes referred to as the adjusted Lie bracket \cite{Compere:2015knw}.} which is designed to take into account the possible field-dependency of the parameters, these vector fields satisfy the commutation relations
\be\label{modified bracket}
\big[\xi(f_1,g_1,h_1,k_1),\xi(f_2,g_2,h_2,k_2)\big]_\star
&=\big[\xi(f_1,g_1,h_1,k_1),\xi(f_2,g_2,h_2,k_2)\big]-\delta_{\xi_1}\xi_2+\delta_{\xi_2}\xi_1\cr
&=\xi(f_{12},g_{12},h_{12},k_{12}),
\ee
where
\bsub\label{commutation relations 3d}
\be
f_{12}&=f_1\partial_uf_2+g_1f_2'-\delta_{\xi_1}f_2-(1\leftrightarrow2),\\
g_{12}&=f_1\partial_ug_2+g_1g'_2-\delta_{\xi_1}g_2-(1\leftrightarrow2),\\
h_{12}&=-\delta_{\xi_1}h_2-(1\leftrightarrow2),\\
k_{12}&=f_1\partial_uk_2+g_1k_2'-\delta_{\xi_1}k_2-(1\leftrightarrow2).
\ee
\esub
In the case where $\delta\xi=0$, the algebra \eqref{commutation relations 3d} is $\big(\text{Diff}(C_2)\loplus\C^\infty(C_2)\big)\oplus\C^\infty(C_2)$, where $C_2$ is the cylinder spanned by $(u,\phi)$ and $\C^\infty(C_2)$ denotes the smooth functions over $C_2$. The functions $f$ and $g$ generate the diffeomorphisms of this cylinder, $k$ generates the translation (with an arbitrary $(u,\phi)$ dependency) of the cylinder along the $r$ direction and $h$ the Weyl rescaling (with an arbitrary $(u,\phi)$ dependency). The action of $h$ on the boundary metric \eqref{boundary metric} is $\delta_h\bar{g}_{\mu\nu}=2h\bar{g}_{\mu\nu}$.

Interestingly, this algebra $\big(\text{Diff}(C_2)\loplus\C^\infty(C_2)\big)\oplus\C^\infty(C_2)$ is reminiscent of the algebra found in \cite{Ciambelli:2021vnn}, however in a very different context and without computing the charges. It would be interesting to study further the relationship between their construction and ours.

Consistently, the subalgebra spanned by $f,g,h$ is precisely the one found in Bondi gauge \cite{ruzziconi2020conservation}. Moreover, the subalgebra spanned by $f,g,k$ is the algebra of residual symmetries found in Fefferman--Graham gauge \cite{Compere:2020lrt,Fiorucci:2020xto,Alessio:2020ioh}. This therefore begs the question of the more precise relationship between the Fefferman--Graham and Bondi gauge at the level of symmetries and later on at the level of the charges, and also raises the question of whether the Bondi--Weyl gauge has a Fefferman--Graham counterpart.

\section{Renormalization, corner ambiguities, and integrable charges}\label{sec:renormalization}

We now address the important issue of symplectic renormalization (and the choice) of the on-shell action and of the potential. This is crucial in order to obtain finite and integrable charges, and will rely on the notion of corner ambiguities. The former requirement guarantees that the observables characterizing a state in the phase space are finite. The latter requirement is the integrability of the charges. It is motivated by the fact that there are no local degrees of freedom in three dimensional gravity, and hence according to \cite{Adami:2020ugu} one expects that there exists a slicing, i.e. a choice of the field dependence of the residual symmetries, such that the charges are integrable. We will show that in Bondi--Weyl gauge integrability requires a corner term in addition to a choice of slicing.

In most treatments of the covariant phase space, the symplectic potential is considered as being ambiguous up to the addition of a total variation and a total exterior derivative \cite{Iyer:1994ys}, i.e. one can shift
\be\label{symppotamb}
\theta\mapsto\theta+\delta b+\de c.
\ee
This follows directly from the fact that the symplectic potential is identified via the variational formula for the Lagrangian, which is $\delta L=\text{EOM}\wedge\delta\Phi+\de\theta$. The ambiguity $b$ can be understood as arising from a shift of the bulk Lagrangian by a boundary term $b=L_{\partial M}$, which is typically done in order to enforce a particular variational principle (e.g. using the Gibbons--Hawking--York term for Dirichlet boundary conditions). The ambiguity $c$, often referred to as corner ambiguity, stems from the fact that $\theta$ is identified as a boundary term in the variation of the Lagrangian.

The important question is of course whether these ambiguities have a physical consequence and meaning. The usual viewpoint on this issue is the following. First, since $\delta^2=0$, when passing from the symplectic potential to the current $\delta\theta$ the $b$-ambiguity drops. This means that although the $b$ term has a physical meaning when discussing the boundary conditions and the variational principle, it does not affect the charges of the theory since these are computed from the $b$-independent symplectic structure. In the case of $c$ however, the shift of the potential survives the passing to the symplectic structure, which therefore acquires a boundary term $\delta c$. This will in turn potentially affect the surface charges, since they are now derived from the symplectic current $\delta\theta+\de(\delta c)$.

As suggested by several authors, there is a natural way in which one can try to relate the $b$ and $c$ ambiguities \cite{Compere:2008us,Harlow:2019yfa,Geiller:2019bti,Freidel:2020xyx,Freidel:2020svx,Freidel:2021cbc}. The way to do so is to realize that typical boundary Lagrangians $b=L_{\partial M}$ have kinetic terms, and can therefore be varied and integrated by parts to isolate their own symplectic potential. It is then natural to treat this co-dimension two symplectic potential of the boundary Lagrangian as the corner contribution $c$ to the total symplectic potential. In this way, the $b$ and $c$ ambiguities are lifted, and these terms are no longer ``ambiguous'', but rather in one-to-one correspondence with the choice of a bulk + boundary Lagrangian defining the theory. This mechanism explains for example the relationship between the Brown--York and Komar charges: they differ by a relative corner charge, which is inherited from the corner potential arising from the Gibbons--Hawking--York boundary Lagrangian. Consistently, this latter is the boundary Lagrangian which relates the Einstein--Hilbert and ADM actions from which the Komar and Brown--York charges are respectively derived.

An important subtlety is that while the proposal of \cite{Compere:2008us,Harlow:2019yfa,Geiller:2019bti,Freidel:2020xyx,Freidel:2020svx,Freidel:2021cbc} relies on examples where a corner contribution $c$ can be derived from the choice of a boundary Lagrangian $b$, it does not explain in general how to reconstruct $b$ from the knowledge of $c$. In particular, in typical situations one can face the need to introduce a corner term $c$ by hand and in the form of a non-covariant component expression (see e.g. \eqref{ren corner} and \eqref{corners} below). In this case, the relationship between $b$ and $c$ is a priori lost and one is forced to face these terms once again as ambiguities. This is precisely what happens in the present study of the on-shell action and charges in Bondi--Weyl gauge. As we are about to explain in details, there are two natural requirements which one can put on $b$ and $c$ (at least to start with). The first one is that $b$ renormalizes the on-shell action (both in $r$ and in $\ell^2$), and the second one that $c$ renormalizes the symplectic potential. This will ensure that both the on-shell action and the charges are finite in the asymptotic limit and in the flat limit. In addition, we will see that another corner term $c$ is needed in order to obtain integrable charges. We will explain in section \ref{sec:origin} below the geometrical origin of these corner terms needed for renormalization and integrability. In particular, the relative corner potentials between metric and tetrad gravity play an important role in this understanding. The clear relationship between these corner terms and the boundary Lagrangian used for renormalization is however still missing, in particular because the boundary Lagrangian is a priori non-unique, and related to a variational principle which we leave unspecified (since the boundary metric is allowed to arbitrarily fluctuate). In this sense, we are therefore forced to interpret $b$ and $c$ as true ambiguities, which we are fixing with physical requirements of finiteness and integrability, but whose covariant geometrical origin remains partly elusive. We note however that in Fefferman--Graham gauge the Comp\`ere--Marolf prescription correctly relates the boundary Lagrangians and corner terms used for renormalization \cite{Compere:2008us,Compere:2020lrt,Fiorucci:2020xto}. It seems therefore that the additional ambiguities encountered in the present work are due to the Bondi gauge. Putting these interpretational issues aside, we now turn to the actual calculations.

\subsection{Renormalized action}
\label{sec:renormalized action}

We first start by discussing, following \cite{ruzziconi2020conservation}, the renormalization of the on-shell action. In order to do so we need to introduce some geometrical quantities and build boundary Lagrangians out of them. We first consider the unit normal to the time-like boundary $\partial M$ at fixed $r$ and the induced metric, given by
\be
n_\mu=\f{1}{\sqrt{g^{rr}}}\delta^r_\mu,
\q\q
n^\mu n_\mu=1,
\q\q
\gamma_{\mu\nu}=g_{\mu\nu}-n_\mu n_\nu.
\ee
We then consider a vector $v$ defined by
\be\label{v vector}
v^\mu\partial_\mu=\f{r}{\sqrt{-\gamma}}\sqrt{\W}(\partial_u+\U\partial_\varphi),
\q\q
v^\mu v_\mu=-1,
\q\q
v^\mu n_\mu=0.
\ee
This is the future-pointing unit vector indicating the direction in which the metric will degenerate in the flat limit $\ell\to\infty$. Equipped with these two vectors, we can then define the boundary Lagrangians
\be
L_\text{GHY}=-\nabla_\mu n^\mu,
\q
L_v=-\nabla_\mu v^\mu,
\q
L_0=-D_\mu v^\mu,
\q
L_\text{ct}=\f{1}{\ell},
\q
L_b=\f{\ell}{2}(D_\mu v^\mu)^2,
\ee
where $D_\mu=\gamma^\alpha_\mu\nabla_\alpha$ is the derivative induced on $\partial M$. Finally, we consider the one-parameter family of covariant boundary Lagrangians\footnote{Note that since
\be
K=\f{2}{\ell}+\f{\ell}{2}R^{(2)}[\gamma]+\O(r^{-3}),
\ee
where $R^{(2)}[\gamma]$ is the Ricci scalar the of two-dimensional boundary metric, we can recombine some factors of $L_\text{GHY}$ with factors of $L_\text{ct}$ if we introduce $R^{(2)}$ as a boundary Lagrangian.}
\be\label{boundary Lagrangian}
L_{\partial M}=(1-\alpha)(L_\text{GHY}+L_v)+\alpha L_0+(3-2\alpha)(L_\text{ct}+L_b),
\ee
where $\alpha$ is a free parameter.

The statement is then that we have a one-parameter family of bulk + boundary actions which is on-shell finite as $r\to\infty$ and $\ell\to\infty$, and given by
\be\label{total action}
S=S_\text{EH}+S_{\partial M}=\f{1}{2}\int_M\sqrt{-g}\,\left(R+\f{2}{\ell^2}\right)+\int_{\partial M}\sqrt{-\gamma}\,L_{\partial M}.
\ee
Indeed, on-shell the bulk action evaluates to
\be\label{on-shell bulk action}
S_\text{EH}\approx-\f{2}{\ell^2}\int_Me^{2\beta+\varphi}(r-H),
\ee
while the boundary action gives
\be\label{on-shell boundary action}
S_{\partial M}&\approx\f{1}{\ell^2}\int_{\partial M}e^{2\beta+\varphi}(r-H)^2\cr
&\phantom{=\f{1}{\ell^2}\int_{\partial M}}-\ell^2e^\varphi\Big(\bar{M}+(1-\alpha)(UH'+\partial_uH)+2\alpha e^{2(\beta-\varphi)}\big(2(\beta')^2+\beta''-\beta'\varphi'\big)\Big)+\O(r^{-1}),\q\q
\ee
where $\bar{M}$ is given in \eqref{M bar}. We see that the on-shell bulk + boundary action is indeed finite as $r\to\infty$ and also in the flat limit $\ell\to\infty$. We note that, although $H$ does not enter in the rescaled boundary metric \eqref{boundary metric}, it enters in the on-shell boundary action and the variational principle.

It is important to note the residual ambiguity of the construction at this stage. Indeed, there might be other boundary Lagrangians, outside of the family \eqref{boundary Lagrangian}, which renormalize the on-shell action in $r$ and $\ell$. Here we have simply exhibited one possible family, where different representatives differ via $\alpha$ at finite and subleading order (i.e. on the second line in \eqref{on-shell boundary action}). It is however natural to have this extra freedom since these finite and subleading terms are related to the choice of boundary conditions and variational principle, which here we have left unspecified. The important message of this subsection is that, because of this remaining freedom in the choice of boundary Lagrangian, there is no unique way to determine corner terms $c$ following the proposal of \cite{Harlow:2019yfa,Geiller:2019bti,Freidel:2020xyx,Freidel:2020svx,Freidel:2021cbc}. Instead, in the following two sections we are going to determine the corner terms $c$ based on criteria of renormalization and integrability. We will only comment in section \ref{sec:origin} on the possible Lagrangian origin of these corner terms.

Let us however point out that a reasonable criterion to further constrain the boundary Lagrangian is to ask that the action be stationary when imposing Dirichlet boundary conditions, which correspond to freezing the boundary metric \eqref{boundary metric} by setting e.g. $\beta=\varphi=U=0$. We show in section \ref{sec:Dirichlet} that $\delta S\approx0$ is achieved with Dirichlet boundary conditions on $\I^+$ if $\alpha=1$.

\subsection{Renormalized potential}

We now proceed with the renormalization of the symplectic potential. This latter, when computed from the bulk Einstein--Hilbert action \eqref{total action}, is given by
\be
\theta_\text{EH}^\mu=\f{1}{2}\sqrt{-g}\big(g^{\alpha\beta}\delta\Gamma^\mu_{\alpha\beta}-g^{\alpha\mu}\delta\Gamma^\beta_{\alpha\beta}\big),
\ee
where we are defining the variations as $\delta g^{\mu\nu}=g^{\mu\alpha}g^{\nu\beta}\delta(g_{\alpha\beta})$. On-shell of \eqref{metric}, the $r$ component of this potential is found to be of the form
\be\label{EH potential r}
\theta_\text{EH}^r\approx r\partial_u(e^\varphi\delta\beta-\delta e^\varphi)+r\left(e^\varphi U\delta\beta-\delta(e^\varphi U)-\f{1}{2}e^\varphi\delta U\right)'-\f{r}{\ell^2}\delta\big(e^{2\beta+\varphi}(r-2H)\big)+\O(r^0).
\ee
We notice that the divergent pieces of this expression as $r\to\infty$ are total variations as well as total angular and $u$ derivatives. First, one can see that the total variations are renormalized by the variation of the first term in the on-shell boundary action \eqref{on-shell boundary action}.\footnote{Although, at the end of the day, what we really care about is having a finite symplectic current $\delta\theta$, so the divergent total variations in \eqref{EH potential r} will drop anyways since $\delta^2=0$ and do not need to be renormalized.} Second, the derivative along the compact coordinate $\phi$ does simply not contribute after integration on the celestial circle where the charges will end up living. Finally, the total $u$-derivative is the term which we have to focus on for the symplectic renormalization. It tells us that we have to introduce the renormalization corner potential
\be\label{ren corner}
\vartheta_\text{ren}=re^\varphi\delta\beta.
\ee
This will of course have an impact on the charges, which will be renormalized thanks to the addition of the charge coming from this corner potential $\vartheta_\text{ren}$.

The contribution to the charge coming from this corner is obtained by using the covariant phase space contraction of $\delta\vartheta_\text{ren}$ with a diffeomorphism transformation, which produces the charge contribution
\be
\slashed{\delta}\Q_\text{ren}=-\oint_S\pounds_\xi\ipp(\delta\vartheta_\text{ren})=-r\oint_Se^\varphi(\delta_\xi\varphi\delta\beta-\delta\varphi\delta_\xi\beta).
\ee
It is this renormalization charge which, when added to the bare charge, will cancel the term of order $r$ and allow to have a finite charge at $r\to\infty$. In fact we will even obtain a stronger result, namely symplectic charges which are completely $r$-independent.

\subsection{Corner ambiguities}

It turns out that, in addition to the corner term used for the renormalization, there are two other corner terms which play an important role in the construction of the charges. First of all, an additional corner potential $\vartheta_1$ is necessary in order to obtain integrable charges. This is similar to what happens in Topologically Massive Gravity for a finite null boundary \cite{Adami:2021sko}. Second, there is another natural corner potential $\vartheta_2$ which controls the Weyl part of the charges and the associated Heisenberg algebra.\footnote{At the end of the day this corner $\vartheta_2$ will be removed by setting $c_2=0$, but we include it in order to illustrate how the Heisenberg sector can be added or removed by playing with the corner ambiguity.} In addition to the renormalization corner potential, we therefore consider the corner contribution
\be\label{corners}
c_1\vartheta_1+c_2\vartheta_2=-c_1e^{\varphi}H\delta\beta+c_2\delta e^\varphi H.
\ee
Interestingly, one can see that for the values $c_1=c_2=1$ this is precisely what results from the action on the term $r\partial_u(e^\varphi\delta\beta-\delta e^\varphi)$ in \eqref{EH potential r} of the finite diffeomorphism $r\mapsto\bar{r}-H$ mentioned above \eqref{large diffeo}. In this sense, the corner potentials \eqref{corners} are produced from the action of the diffeomorphism on the renormalization corner term, with the extra subtlety that a term which was previously a total variation now gives rise to a non-vanishing corner symplectic current. As we will see below when computing the charges however, while $c_1=1$ is indeed required for integrability, any value of $c_2$ is allowed, and $c_2=1$ is peculiar in that it kills the Weyl charges and the associated Heisenberg algebra. Keeping for the moment arbitrary values of the corner couplings, the contribution of the corner potentials $\vartheta_1$ and $\vartheta_2$ to the diffeomorphism charge is found to be of the form
\bsub\label{corner charges}
\be
\slashed{\delta}\Q_1&=-\oint_S\pounds_\xi\ipp(\delta\vartheta_1)=\oint_S\delta_\xi(e^\varphi H)\delta\beta-\delta(e^\varphi H)\delta_\xi\beta,\\
\slashed{\delta}\Q_2&=-\oint_S\pounds_\xi\ipp(\delta\vartheta_2)=\oint_Se^\varphi\big(\delta_\xi\varphi\delta H-\delta\varphi\delta_\xi H\big),
\ee
\esub
where the field variations are given in \eqref{variations}.

Putting all of this together, we get that the total diffeomorphism charge with its three corner contributions is
\be\label{total charge}
\slashed{\delta}\Q=\slashed{\delta}\big(\Q_\text{EH}+\Q_\text{ren}+c_1\Q_1+c_2\Q_2\big),
\ee
where
\be
\slashed{\delta}\Q_\text{EH}=\oint_S\sqrt{-g}\,\eps_{\mu\nu\phi}\left(\nabla_\alpha\delta g^{\alpha\nu}-\xi^\mu\big(\nabla^\nu\delta g\big)-\xi_\alpha\nabla^\mu\delta g^{\alpha\nu}-\f{1}{2}\delta g\nabla^\mu\xi^\nu+\delta g^{\mu\alpha}\nabla_\alpha\xi^\nu\right)
\ee
%\be
%\slashed{\delta}\Q_\text{EH,BB}=\f{1}{2}\oint_S\sqrt{-g}\,\eps_{\mu\nu\phi}\left(\xi^\mu\big(\nabla^\nu\delta g-\nabla_\alpha\delta g^{\alpha\nu}\big)+\xi_\alpha\nabla^\mu\delta g^{\alpha\nu}+\f{1}{2}\delta g\nabla^\mu\xi^\nu+\f{1}{2}\delta g^{\mu\sigma}(\nabla^\nu\xi_\sigma-\nabla_\sigma\xi^\nu)\right)
%\ee
is the Iyer--Wald charge \cite{Iyer:1994ys} computed from the Einstein--Hilbert Lagrangian, and defined here with fundamental variations on the lower indices, i.e. $\delta g^{\mu\nu}=g^{\mu\alpha}g^{\nu\beta}\delta(g_{\alpha\beta})$ and $\delta g=g^{\mu\nu}\delta g_{\mu\nu}$, and with the convention $\eps_{ur\phi}=1$. Finally, let us mention already at this stage that the Iyer--Wald and Barnich--Brandt charges \cite{Barnich:2001jy,Barnich:2003xg,Barnich:2007bf} coincide in Bondi--Weyl gauge.

We are now ready to go ahead with the computation of these diffeomorphism charges and of their algebra. This requires the choice of an integrable slicing, as we now explain.

\subsection{Integrable charges}
\label{sec:slicing}

When computed with the vector fields \eqref{AKV} the charges \eqref{total charge} are generically non-integrable, regardless of the value of $c_1$ (which as we will see we have included for further integrability issues). We do not give the explicit expression of this non-integrable charge because it is lengthy and not particularly enlightening. When dealing with non-integrable charges, one possibility is to consider the modified Barnich--Troessaert bracket \cite{Barnich:2011mi}, or other proposals for a modified bracket such as that in \cite{Freidel:2021cbc}. However, as argued in \cite{Adami:2020ugu,ruzziconi2020conservation,Adami:2021sko}, the fact that we are in three-dimensional gravity means that there should be no physical flux sourcing the non-integrability (since there are no degrees of freedom which can leak through the boundary). Indeed, as we are about to see, integrability can be achieved using a combination of two mechanisms:
\begin{itemize}
\item[$i)$] a change of slicing, which amounts to a (non-unique but invertible) field-dependent redefinition of the vector field generators $(f,g,h,k)$,
\item[$ii)$] the introduction of the corner potential \eqref{corners} with $c_1=1$ and $c_2$ arbitrary.
\end{itemize}
Already at this point, let us mention that integrability cannot be achieved only with a change of slicing and without introducing a corner term. This shows that the relationship between integrability and the absence of local degrees of freedom is more subtle than it seems, and requires to properly understand the corner ambiguities.

Achieving integrability amounts to the resolution of a Pfaff problem \cite{Barnich:2007bf,Adami:2020ugu,ruzziconi2020conservation}. The first ingredient here is a change of the vector field generators, from $(f,g,h,k)$ to field-independent functions $(\tilde{f},\tilde{g},\tilde{h},\tilde{k})$ via the change of slicing\footnote{The choice of an integrable slicing is not unique, see \cite{Adami:2020ugu} for a discussion.}
\bsub\label{slicing}
\be
f&=\tilde{f}e^{\varphi-2\beta},\\
g&=\tilde{g}+\tilde{f}e^{\varphi-2\beta}U,\\
h&=-\big(\tilde{h}+(\tilde{g}e^\varphi)'\big)e^{-\varphi}+\tilde{g}'+\tilde{g}\varphi',\\
k&=\tilde{k}-\tilde{g}H'-\tilde{f}e^{\varphi-(c_1+1)\beta}(UH'+\partial_uH).
\ee
\esub
Using this, and freely integrating by parts on $\phi$, we find that the diffeomorphism charge \eqref{total charge} takes the form
\be\label{generic charge}
\slashed{\delta}\Q=\oint_S&\ \tilde{f}\delta\tilde{M}+\tilde{g}\delta\tilde{N}+(1-c_2)\tilde{h}\delta H+(1-c_2)\tilde{k}\delta e^\varphi\cr
&+\f{1}{2}(c_1-1)\left[\tilde{h}\big(H\delta(\varphi-2\beta)+\delta H\big)+2\tilde{k}e^\varphi\delta\beta-\tilde{g}\left(2\delta(e^\varphi H)\beta'+e^\varphi\big[(H\delta\varphi)'+\delta H'\big]\right)\right]\cr
&-\f{1}{2}(c_1-1)\partial_u\tilde{f}e^{2(\varphi-\beta)}(\delta H+H\delta\varphi)\cr
&+\f{1}{2}(c_1-1)\tilde{f}e^{2(\varphi-\beta)}\Big[\big(2U'+2U(\varphi-\beta)'+U\partial_\phi\big)(\delta H+H\delta\varphi)+2\delta\beta(UH'+\partial_uH)\Big]\cr
&-\tilde{f}e^{(3-c_2)\varphi-2\beta}(UH'+\partial_uH)\delta\Big[e^{(c_2-1)\varphi}\Big(1-e^{(1-c_1)\beta}\Big)\Big],
\ee
where in the new slicing the Bondi mass and angular momentum aspects are given by\footnote{We note that in terms of $\Phi$ defined by $\Phi'=e^{\varphi-2\beta}$ we can rewrite $\tilde{M}=-\text{Sch}[\Phi]+2(\beta')^2+e^{2(\varphi-\beta)}\bar{M}$, where $\text{Sch}[\Phi]=\big(\Phi'\Phi'''-\f{3}{2}(\Phi'')^2\big)/(\Phi')^2$ is the Schwarzian derivative.}
\bsub
\be
\tilde{M}&=4(\beta')^2-2\beta'\varphi'+\f{1}{2}(\varphi')^2+(2\beta-\varphi)''+e^{2(\varphi-\beta)}\bar{M},\\
\bar{M}&=M-(UH)'-\partial_uH-H(U\varphi'+\partial_u\varphi)+\f{e^{2\beta}}{2\ell^2}H^2,\\
\tilde{N}&=e^\varphi\bar{N}=e^\varphi\big(N+2H\beta'-H'\big).
\ee
\esub
Note that these charges are symplectic, i.e. they do not depend on $r$. We have written the charge in a way which makes explicit the fact that integrability can be achieved with $c_1=1$. In this case, we get
\be\label{final charge}
\Q=\oint_S\tilde{f}\tilde{M}+\tilde{g}\tilde{N}+(1-c_2)\tilde{h}H+(1-c_2)\tilde{k}e^\varphi.
\ee
This calculation also makes it clear that the addition of the corner potential $\vartheta_2$ with coupling $c_2$ simply shifts the charges carried by $\tilde{h}$ and $\tilde{k}$. While in the general case the natural value is $c_2=0$ since there is no physical reason to actually include this term (at the difference with $\vartheta_\text{ren}$ and $\vartheta_1$), this shows that the charges added by the extension to Bondi--Weyl gauge can be removed with a corner term. This indicates that the so-called ambiguities in the choice of the symplectic potential \eqref{symppotamb} can carry physical content and hence should carefully be studied when considering the charges.

The charges associated to the symmetry generators $(\tilde{f},\tilde{g},\tilde{h},\tilde{k})$ are generically non-vanishing. These symmetries are therefore large and the phase space carries four unconstrained time-dependent charges. Until now, the maximal number of charges that were found in the metric formulation was three \cite{Adami:2020ugu,Adami:2021sko}. In these works however, the boundary was taken at fixed $r$. Here we see that this condition is too restrictive because the charge associated with $\tilde{k}$ can only exist if the boundary is allowed to move. Moreover, since our charges are symplectic, the analysis leading to the four charges is indeed valid for any finite boundary.

Just like in the case of the general Bondi gauge \cite{ruzziconi2020conservation}, these charges are not conserved (none of the four generators) in spite of being integrable. This was expected as we allow unconstrained boundary sources and hence do not require $\delta S\approx0$.

Using the new slicing in the asymptotic Killing vector, we can look once again at the action on the metric and deduce that the fields transform as
\bsub\label{transformations in new slicing}
\be
2\delta_{\tilde\xi}\beta&=\tilde{g}(2\beta-\varphi)'-\tilde{g}'-e^{-\varphi}\tilde{h}+e^{-(2\beta-\varphi)}(U\tilde{f}'-\tilde{f} U'+\partial_u\tilde{f}),\\
\delta_{\tilde\xi}e^\varphi&=-\tilde{h},\\
\delta_{\tilde\xi}U&=\tilde{g} U'-U\tilde{g}'-\partial_u \tilde{g}+\f{e^{2\beta-\varphi}}{\ell^2}\left(\tilde{f}'-\tilde{f}(2\beta-\varphi)'\right),\\
\delta_{\tilde\xi}H&=\tilde{k}.
\ee
\esub
With this, one can easily see for example how the corner term $\vartheta_2$ contributes to the charge as
\be
-\pounds_{\tilde{\xi}}\ipp(\delta\vartheta_2)=\pounds_{\tilde{\xi}}\ipp(\delta e^\varphi\delta H)=\delta_{\tilde{\xi}}e^\varphi\delta H-\delta e^\varphi\delta_{\tilde{\xi}}H=-\tilde{h}\delta H-\tilde{k}\delta e^\varphi.
\ee

With this new slicing we can use once again the definition \eqref{modified bracket} of the adjusted Lie bracket, where now the field-dependent contributions play a crucial role, to find
\be\label{modified bracket tilde}
\big[\tilde{\xi}(\tilde{f}_1,\tilde{g}_1,\tilde{h}_1,\tilde{k}_1),\tilde{\xi}(\tilde{f}_2,\tilde{g}_2,\tilde{h}_2,\tilde{k}_2)\big]_\star=\tilde{\xi}(\tilde{f}_{12},\tilde{g}_{12},\tilde{h}_{12},\tilde{k}_{12}),
\ee
with
\bsub\label{commutation relations new slicing}
\be
\tilde{f}_{12}&=\tilde{f}_1\tilde{g}_2'+\tilde{g}_1\tilde{f}_2'-(1\leftrightarrow2),\\
\tilde{g}_{12}&=\tilde{g}_1\tilde{g}'_2+\f{1}{\ell^2}\tilde{f}_1\tilde{f}_2'-(1\leftrightarrow2),\\
\tilde{h}_{12}&=0\,,\\
\tilde{k}_{12}&=0\,.
\ee
\esub
This is an algebroid where the base space is parametrized by $u$. Using $\tilde f_\pm=\tilde g\pm \tilde f/\ell$ for non vanishing $\ell$, the explicit form of this algebroid is $\big[\text{Diff}(S_1)\oplus \text{Diff}(S_1)\oplus\C^\infty(S_1)\oplus\C^\infty(S_1)\big]_u$,  and in the flat limit one has $\big[\big(\text{Diff}(S_1)\loplus \text{Vect}(S_1)\big)\oplus\C^\infty(S_1)\oplus\C^\infty(S_1)\big]_u$, where in the first factor we recognize $\text{BMS}_3=\text{Diff}(S_1)\loplus \text{Vect}(S_1)$. By comparing with \eqref{commutation relations 3d} one can see that in the integrable slicing \eqref{slicing} the cosmological constant has reappeared in the commutation relations.

Finally, having integrable charges, we are guaranteed by the representation theorem that the phase space bracket, which is defined as
\be
\big\{\Q[\tilde{\xi}_1],\Q[\tilde{\xi}_2]\big\}=-\delta_{\tilde{\xi}_1}\Q[\tilde{\xi}_2],
\ee
reproduces the modified bracket \eqref{modified bracket tilde} up to central extensions. We indeed find that
\be
\big\{\Q[\tilde{\xi}_1],\Q[\tilde{\xi}_2]\big\}=\Q[\tilde{\xi}_1,\tilde{\xi}_2]_\star+\oint_S\tilde{f}_1\tilde{g}_2'''-\tilde{f}_2\tilde{g}_1'''+(c_2-1)\oint_S\tilde{h}_1\tilde{k}_2-\tilde{h}_2\tilde{k}_1,
\ee
which reveals gravitational central extensions for $\mathfrak{vir}\oplus\mathfrak{vir}$. In addition, we see that the Weyl sector also receives a central extension, and therefore becomes an Heisenberg algebra. The total algebra after the change of slicing is therefore $\mathfrak{vir}\oplus\mathfrak{vir}\oplus\text{Heisenberg}$.

When removing the field $H$ the Heisenberg part drops out and we are back to the general Bondi gauge \cite{ruzziconi2020conservation}. The algebra then reduces to an algebroid with base space parametrized by $u$, and the algebra at each $u$ consists in two copies of Virasoro with Brown--Henneaux central charges \cite{Brown:1986nw}, and to the BMS$_3$ algebra \cite{Barnich:2006av} in the flat limit $\ell\to\infty$. In this case the charges are still non-conserved.

In the next section we further reduce to Dirichlet boundary conditions by removing all the fields but $M$, $N$ and $H$ (i.e. freezing the boundary metric). The $u$-dependency then drops and we obtain conserved charges.

\subsection{Dirichlet boundary conditions}
\label{sec:Dirichlet}

We now briefly discuss the limit to Dirichlet boundary conditions, where the boundary metric \eqref{boundary metric} is fixed. The simplest choice is to set $\beta=\varphi=U=0$. Then we still have the function $H$ in our solution space. Let us first look at the symplectic potential and the on-shell variation of the action.

Using these Dirichlet boundary conditions, the on-shell symplectic potential has a radial component \eqref{EH potential r} which reduces to
\be
\theta_\text{EH}^r\approx\delta\left(\bar{M}-\f{H^2}{\ell^2}+r\f{2H}{\ell^2}\right)+\O(r^{-1}),
\ee
while the on-shell boundary action \eqref{on-shell boundary action} reduces to
\be
S_{\partial M}&\approx\f{1}{\ell^2}\int_{\partial M}(r-H)^2-\ell^2\Big(\bar{M}+(1-\alpha)\partial_uH)\Big)+\O(r^{-1}).
\ee
From this we can see that $\delta S\approx0$ can be achieved with Dirichlet boundary conditions provided we set $\alpha=1$ in the boundary Lagrangian (although $\alpha$ can be arbitrary if $H=0$). This is consistent with \cite{ruzziconi2020conservation}, and simply generalizes their result to $H\neq0$.

Since we have set fields to zero for Dirichlet boundary conditions, looking at \eqref{transformations in new slicing} puts constraints on the parameters of the asymptotic Killing vector field. We have
\bsub\label{constraints f g}
\be
\beta=0\ &\Rightarrow\ \partial_uf=g',\\
U=0\ &\Rightarrow\ \partial_ug=\f{f'}{\ell^2},\\
\varphi=0\ &\Rightarrow\ h=0,
\ee
\esub
where now the change of slicing $(\tilde{f},\tilde{g},\tilde{h})=(f,g,h)$ is trivial. The mass and angular momentum aspect reduce to
\be
\tilde{M}=\bar{M}=M-\partial_uH+\frac{H^2}{2\ell^2},
\q\q
\tilde{N}=\bar{N}=N-H',
\ee
and satisfy the evolution equations
\be\label{evolution Dirichlet}
\partial_u\tilde{M}=\f{\tilde{N}'}{\ell^2},
\q\q
\partial_u\tilde{N}=\tilde{M}'.
\ee
%Their transformation law is
%\bsub
%\be
%\delta_\xi\tilde{M}&=2\tilde{M}g'+\tilde{M}'g+\f{1}{\ell^2}\big(2\tilde{N}f'+\tilde{N}'f\big)-g''',\\
%\delta_\xi\tilde{N}&=2\tilde{M}f'+\tilde{M}'f+2\tilde{N}g'+\tilde{N}'g-f'''.
%\ee
%\esub
Finally, the charge reduces to\footnote{We either derive this from \eqref{generic charge} using the fact that $\tilde{k}\delta e^\varphi|_{\varphi=0}=\tilde{k}\delta(1)=0$, or from \eqref{final charge} treating $\tilde{k}$ as a field-independent integration constant.}
\be\label{Dirichlet charges}
\Q=\oint_Sf\tilde{M}+g\tilde{N}.
\ee
We therefore see that the symmetry generator $k$ does not appear in the Dirichlet charge, meaning that in this case it is associated with a true gauge freedom. We can then use this freedom to arbitrarily fix the value of $H$, for example to $H=0$ to recover the results of \cite{ruzziconi2020conservation}.

Finally, we can use the constraints \eqref{constraints f g} and the evolution equations \eqref{evolution Dirichlet} to show that the charge is conserved, i.e. $\partial_u\Q=0$.

\section{Triad formulation}
\label{sec:triad}

We now make a detour through the triad formulation of gravity. Although at the end of the day the charges derived in triad variables agree, as they should, with \eqref{final charge} derived in the metric formulation, the precise proof of this result contains many subtleties which turn out to be related to anomalies and corner ambiguities. We will show in particular how the corner term $\vartheta_\text{ren}+\vartheta_1$ naturally appears in triad gravity. From now on we will take $c_2=0$ since this is the canonical value.

We recall that the triad formulation uses Lie algebra-valued one-forms $e^i_\mu$, out of which the metric is constructed as $g_{\mu\nu}=e^i_\mu e^j_\nu\eta_{ij}$. Denoting the triad and metric Lagrangians by Einstein--Cartan and Einstein--Hilbert respectively, we have\footnote{With respect to the general Lagrangian studied in \cite{Geiller:2020edh}, in $L_\text{EC}$ we take $\sigma_0=1/(2\ell^2)$, $\sigma_1=1/2$, and $\sigma_2=\sigma_3=0$. This implies $p=-\Lambda=1/\ell^2$ and $q=0$. With respect to \cite{ruzziconi2020conservation} this corresponds to $8\pi G=1$.}
\be\label{1 and 2 order Lagrangians}
L_\text{EC}=e\wedge\left(F+\f{1}{6\ell^2}[e\wedge e]\right),
\q\q
L_\text{EH}=\f{1}{2}\sqrt{-g}\,\left(R+\f{2}{\ell^2}\right).
\ee
Here $F=\de\omega+[\omega\wedge\omega]/2$ is the curvature of the gauge connection $\omega^i_\mu$, and $[p\wedge q]^i={\eps^i}_{jk}p^j\wedge q^k$ denotes the Lie algebra commutator. The connection can be taken as an independent variable in the first order formulation. On-shell of the torsion equation of motion obtained by varying with respect to $\omega$, we have $\omega\simeq\omega(e)$, and we get of course that $L_\text{EC}\simeq L_\text{EH}$. In particular, when going completely on-shell \eqref{on-shell bulk action} comes from
\be\label{on-shell equivalence EC EH}
L_\text{EC}\approx-\f{1}{3\ell^2}e\wedge[e\wedge e]=-\f{2}{\ell^2}\sqrt{-g}\approx L_\text{EH}.
\ee
So far, this expresses the fact that the two formulations are equivalent at the level of the Lagrangian and of the equations of motion. Importantly, this equivalence does however not extend to the symplectic level, where the potentials differ. Explicitly, they are given by
\be
\theta_\text{EC}^\mu=\f{1}{2}\eps^{\mu\nu\rho}(\delta\omega\wedge e)_{\nu\rho},
\q\q
\theta_\text{EH}^\mu=\f{1}{2}\sqrt{-g}\big(g^{\alpha\beta}\delta\Gamma^\mu_{\alpha\beta}-g^{\alpha\mu}\delta\Gamma^\beta_{\alpha\beta}\big),
\ee
where $\eps^{\mu\nu\rho}$ is the Levi--Civita symbol and $\eps^{r\phi u}=1$, and we have dualized the two-form $\theta_\text{EC}=\delta\omega\wedge e$ so that it has a vector index. The statement is then that we have $\theta_\text{EC}\neq\theta_\text{EH}$ \cite{DePaoli:2018erh,Oliveri:2019gvm,Oliveri:2020xls,Freidel:2020xyx,Freidel:2020svx}. A lot can be learned by looking at how exactly the potentials differ, and how this difference can be fixed so as to derive the same charges in both formulations.

In section \ref{subsec:triad potential} we explicitly evaluate the potential in triad variables to show that it differs from the metric potential, and explain how this modifies the procedure of symplectic renormalization.

For completeness we study in section \ref{subsec:improved} the symmetries of the triad. This latter transforms under both internal Lorentz transformations and diffeomorphisms. When looking at asymptotic symmetries we have to enhance the diffeomorphisms by adding to them a specific Lorentz transformation. This amounts in fact to working with the field-space Lie derivative $\mathsf{L}_\xi$ instead of the spacetime derivative $\pounds_\xi$, which differs from the former by a term which can be interpreted as an anomaly  \cite{Hopfmuller:2018fni,Chandrasekaran:2020wwn,Freidel:2021cbc}.\footnote{We stress that the triad formulation is no anomalous per se, and that the computation of the charges is non-ambiguous if we properly define the asymptotic symmetries in terms of diffeomorphisms and Lorentz transformations, which are the symmetries of $e^i_\mu$. However, if we insist on comparing the triad formulation with the metric one, we can then interpret the fact that $\delta_\xi e\neq\pounds_\xi e$ as an anomaly.}

We finally explain in section \ref{subsec:DPS} how to resolve the symplectic mismatch between the triad and metric formulations, and how to derive the same diffeomorphism charges in both cases. The way to do so is to force the triad and metric potentials to agree by adding a corner term to the triad potential \cite{DePaoli:2018erh,Oliveri:2019gvm,Oliveri:2020xls}. In the language of \cite{Freidel:2020xyx,Freidel:2020svx}, this is the relative corner term $\vartheta_\text{EC/EH}$ between the EC and EH formulations.

Finally, in section \ref{subsec:dualcharges} we investigate the charges associated to internal symmetries (i.e. without referring to diffeomorphims). We show that at the end of the day there is exactly the same amount of information in the metric and triad formalism in Bondi--Weyl gauge, namely four independent co-dimension one large symmetry generators.

\subsection{Symplectic potential}
\label{subsec:triad potential}

To write down the potential in triad variables, we choose the internal metric and the triad as
\be\label{triad}
\eta_{ij}=
\begin{pmatrix}
0&1&0\\
1&0&0\\
0&0&1
\end{pmatrix},
\q\q
\everymath={\displaystyle}
e^i_\mu=
\begin{pmatrix}
\f{1}{2}\left(\f{\V}{r}\B+r^2\U^2\W\right)&1&0\\
-\B&0&0\\
-r^2\U\W&0&r\sqrt{\W}
\end{pmatrix}
=
\begin{pmatrix}
e^0_u&e^1_u&0\\
e^0_r&0&0\\
e^0_\phi&0&e^2_\phi
\end{pmatrix}.
\ee
To compare the potential with the metric expression \eqref{EH potential r} we have to go on-shell of the torsion and express $\omega$ in terms of $e$ as
\be
\omega^i_\mu(e)=\f{1}{2}{\eps^i}_{jk}\hat{e}^{j\alpha}\nabla_\mu e^k_\alpha,
\ee
where $\hat{e}^{i\alpha}=g^{\alpha\beta}e^i_\beta$ is the inverse triad. From now on it will be understood that $\omega=\omega(e)$. The component of interest in the dualized symplectic potential is the radial one, which reads
%\bsub
%\be
%\theta_\text{EC}^u&=e^\varphi\delta(\varphi-2\beta),\\
%\theta_\text{EC}^r
%&=\delta(e^\varphi M)+e^\varphi M\delta(\varphi-2\beta)+e^\varphi N\delta U+2e^{2\beta-\varphi}\delta(\beta')^2-\delta\partial_u(e^\varphi H)\cr
%&\pe+e^\varphi H\left(\delta\partial_u(\varphi-2\beta)+\delta U'+\delta U(\varphi+2\beta)'+U\delta(\varphi-2\beta)'-\f{1}{\ell^2}\delta(e^{2\beta}H)\right)+\O(r),\label{theta phi u}\\
%\theta_\text{EC}^\phi&=2e^\varphi\big(\delta U+U\delta(\varphi-\beta)\big),
%\ee
%\esub
\be\label{EC potential r}
\theta_\text{EC}^r=2r\big(e^\varphi U\delta\beta-\delta(e^\varphi U)\big)'+r\partial_u(2e^\varphi\delta\beta-\delta e^\varphi)+\f{r}{\ell^2}\delta\big(e^{2\beta+\varphi}(2H-r)\big)+\O(r^0).
\ee
This should then be compared to \eqref{EH potential r}. In particular, we see that the corner term which matters for the renormalisation of the charges is now
\be
2\vartheta_\text{ren}=2re^\varphi\delta\beta,
\ee
i.e. twice the corner term needed in the metric formulation. This shows that even if the triad and metric formulations agree on-shell in the sense \eqref{on-shell equivalence EC EH}, and are in particular renormalized in the same way at the Lagrangian level, they are not renormalized in the same way at the level of the symplectic potential and of the charges.

\subsection{Improved diffeomorphisms}
\label{subsec:improved}

Diffeomorphisms act on the triad, seen as a one-form, via the Lie derivative $\pounds_\xi=\de(\xi\ip\,\cdot)+\xi\ip(\de\,\cdot)$. However, one can explicitly check that, with the vector field \eqref{AKV} (using the vector field in the new slicing does not work either), this Lie derivative does not preserve the triad \eqref{triad} , i.e. we have $\pounds_\xi\big(e(\Phi)\big)\neq e(\delta_\xi\Phi)$.

To obtain the proper transformation law, we have to improve the diffeomorphism of the triad by infinitesimal Lorentz transformations. Recall that Lorentz transformations act on the fields as $\delta^\text{L}_\alpha e=[e,\alpha]$ and $\delta^\text{L}_\alpha\omega=\de_\omega\alpha=\de\alpha+[\omega,\alpha]$, and give a charge
\be
\slashed{\delta}\J_\alpha=-\delta^\text{L}_\alpha\ipp\Omega_\text{EC}=\oint_S\alpha\delta e.
\ee
In the present case, the transformations required in order to improve the diffeomorphisms can be parametrized by two Lie algebra elements $\rho^i$ and $\lambda^i$. The first one is given by
\be
\rho^i=-\f{1}{2}{\eps^i}_{jk}\hat{e}^{j\mu}\pounds_\xi e^k_\mu,
\ee
and defines the so-called Kosmann derivative
\be
\mathscr{K}_\xi e=\pounds_\xi e+\delta^\text{L}_\rho e=\pounds_\xi e+[e,\rho],
\ee
which as one can check is such that $\mathscr{K}_\xi e=0$ when $\xi$ is Killing \cite{Jacobson:2015uqa,Prabhu:2015vua,DePaoli:2018erh}. The second gauge parameter is defined in terms of the components of the Kosmann derivative acting on $e^i_\mu$ as
\be
\lambda^i=(\lambda^0,\lambda^1,\lambda^2)=\big((\mathscr{K}_\xi e)^2_u,0,-(\mathscr{K}_\xi e)^1_u\big).
\ee
With these two parameters, one can explicitly check that
\be\label{enhanced diffeo on e}
\big(\pounds_\xi+\delta^\text{L}_\rho+\delta^\text{L}_\lambda\big)\big(e(\Phi)\big)=e(\delta_\xi\Phi),
\ee
where the variations are \eqref{variations}. It is therefore only by correcting the Lie derivative with Lorentz transformations that the vector field \eqref{AKV}, which was found by the condition of invariance of the metric, defines a symmetry of the triad as well.

Notice that because $\lambda$ is itself defined in terms of the Kosmann derivative, we have that $\mathscr{K}_\xi$ and $\mathscr{K}_\xi+\delta^\text{L}_\lambda$ both annihilate $e$ when $\xi$ is Killing. There is therefore a residual ambiguity in this definition of the Kosmann derivative, which can only be fixed by looking at the symmetries preserving $e$. We also point out that while the definition of $\rho$ is canonical, $\lambda$ depends on the Lorentz frame which has been chosen for $e$. Here we have been fortunate enough that with the ``simple'' choice of triad \eqref{triad} $\lambda$ has a compact expression. Equivalently, one can of course also forget about the split between $\rho$ and $\lambda$, and look for the symmetry parameters $\xi$ and $\alpha$ such that $\pounds_\xi+\delta^\text{L}_\alpha$ are asymptotic symmetries preserving $e$. Consistently, this leads to the same result, namely that $\xi$ is given by \eqref{AKV} and $\alpha=\rho+\lambda$.

Let us end with a comment on anomalies. Usually one does not refer to the Lorentz transformations in \eqref{enhanced diffeo on e} as anomalies, since they are just features of the symmetry structure of triad gravity. However, if we ground ourselves in the metric formalism studied in the previous sections, where the transformations $\delta_\xi$ of the fields have been determined, we can view the triad $e$ as an object which is anomalous exactly in the sense \eqref{enhanced diffeo on e}. Indeed, using the notation of \cite{Hopfmuller:2018fni,Chandrasekaran:2020wwn,Freidel:2021cbc} we can rewrite this equation as
\be\label{triad anomaly}
\delta_\xi e=(\pounds_\xi+\Delta_\xi)e=\mathsf{L}_\xi e,
\ee
where $\mathsf{L}_\xi=\delta(\delta_\xi\ipp\cdot)+\delta_\xi\ipp(\delta\,\cdot)$ is the field-space Lie derivative. The action of this latter on the triad differs from the spacetime Lie derivative by the anomaly term $\Delta_\xi e=\delta^\text{L}_{\rho+\lambda}e$. In section \ref{sec:origin} we show that this field space Lie derivative $\mathsf{L}_\xi$ can be used in the metric formulation to write down the contribution of the corner charges $\Q_\text{ren}+\Q_1$ as coming from a corner term in triad variables.

\subsection{Relative corner term}
\label{subsec:DPS}

We now explain how one can force the triad and metric formulations to agree at the symplectic level by matching their symplectic potentials. This is done by realizing that the two potentials actually differ by a corner term. This term was originally introduced in \cite{DePaoli:2018erh}, and later studied in \cite{Oliveri:2019gvm,Oliveri:2020xls,Freidel:2020xyx,Freidel:2020svx}. In three spacetime dimensions, this corner term is a one-form which reads
\be\label{DPS corner}
(\vartheta_\text{EC/EH})_\mu=-\f{1}{2}\eps_{ijk}e^i_\mu\hat{e}^{j\alpha}\delta e^k_\alpha,
\ee
and in terms of which we have
\be\label{EH = EC + DPS}
\theta_\text{EH}=\theta_\text{EC}+*\de\vartheta_\text{EC/EH},
\ee
or in components $\theta_\text{EH}^\mu=\theta_\text{EC}^\mu+\eps^{\mu\nu\rho}\partial_\nu(\vartheta_\text{EC/EH})_\rho$. This identity justifies the name EC/EH, as this co-dimension two form is the relative corner term between the Einstein--Hilbert and Einstein--Cartan formulations. One can check that the Lorentz charges arising from this corner term are $-\delta_\alpha\ipp\Omega_\text{EC/EH}=-\J_\alpha$. This implies in turn that the Lorentz charges coming from the extended potential $\theta_\text{EC}+*\de\vartheta_\text{EC/EH}$ vanish, as they should in order to match the metric formulation.

It is now natural to ask what is the contribution of the corner term \eqref{DPS corner} to the diffeomorphism charges associated with the Lie derivative $\pounds_\xi$. For this we compute
\be\label{Lie in DPS}
\slashed{\delta}\Q_\text{EC/EH}=-\pounds_\xi\ipp\Omega_\text{EC/EH}=\f{1}{2}\oint_S\eps_{ijk}\left(\left[\big(\pounds e^i\big)_\phi\hat{e}^{j\alpha}+e^i_\phi\big(\pounds_\xi\hat{e}^j\big)_\alpha\right]\delta e^k_\alpha-\delta\big(e^i_\phi\hat{e}^{j\alpha}\big)\big(\pounds e^k\big)_\alpha\right).
\ee
One can check that, as expected, this has just the effect of relating the EH and the EC charges as
\be\label{EH = EC + DPS for charges}
\slashed{\delta}\Q_\text{EH}=\slashed{\delta}\big(\Q_\text{EC}+\Q_\text{EC/EH}\big),
\ee
where the charge in triad variables is
\be\label{triad charge}
\slashed{\delta}\Q_\text{EC}=-\pounds_\xi\ipp\Omega_\text{EC}=\oint_S(\xi\ip\omega)\delta e+(\xi\ip e)\delta\omega.
\ee
Relation \eqref{EH = EC + DPS for charges} between the charges consistently reflects relation \eqref{EH = EC + DPS} between the potentials.

Note that here we have computed the charges by contracting the spacetime Lie derivative $\pounds_\xi$ with the various symplectic structures. But it is also interesting to do so with the field-space Lie derivative $\mathsf{L}_\xi$ introduced in the previous section. In this previous section we have explained that $\pounds_\xi\ipp\Omega_\text{EC}\neq\mathsf{L}_\xi\ipp\Omega_\text{EC}$. Similarly, when acting on the relative corner term we find
\bsub\label{QDPS}
\be
-\pounds_\xi\ipp\Omega_\text{EC/EH}&=\slashed{\delta}\big(\Q_\text{ren}+\Q_1+\J_\rho+\J_\lambda\big),\\
-\mathsf{L}_\xi\ipp\Omega_\text{EC/EH}&=\slashed{\delta}\big(\Q_\text{ren}+\Q_1\big).
\ee
\esub
Consistently, acting with either $\pounds_\xi$ or $\mathsf{L}_\xi$ on $\Omega_\text{EC}+\Omega_\text{EC/EH}$ gives the same result, since after all this is equal to the action on the covariant symplectic structure $\Omega_\text{EH}$ by virtue of \eqref{EH = EC + DPS}. We see however that depending on which Lie derivative we choose the contribution from the Lorentz charges $\J_\rho+\J_\lambda$ moves around: it is produced by $\Omega_\text{EC/EH}$ when we use $\pounds_\xi$, while it is produced by $\Omega_\text{EC}$ when we use $\mathsf{L}_\xi$. In either case, $\Omega_\text{EC/EH}$ also takes care of bringing the necessary renormalization and integrability contribution $\Q_\text{ren}+\Q_1$. Note that with this the total charge \eqref{total charge} (where now we take $c_2=0$ and $c_1=1$ for integrability) becomes
\be\label{Q total via EC}
\slashed{\delta}\Q
&=\slashed{\delta}\big(\Q_\text{EH}+\Q_\text{ren}+\Q_1\big)\cr
&=\slashed{\delta}\big(\Q_\text{EC}+2\Q_\text{ren}+2\Q_1+\J_\rho+\J_\lambda\big).
\ee

To conclude this section, it is interesting to note that the relative corner term $\vartheta_\text{EC/EH}$ ``knows'' about the renormalization and integrability corner terms, as can be seen on \eqref{QDPS}. We come back to this in section \ref{sec:origin}.

\subsection{Charges from internal gauge transformations}
\label{subsec:dualcharges}

In this section we derive the charges associated with the internal symmetries of the triad. This method is specific to the three-dimensional case, and exploits the topological nature of the theory to trade the diffeomorphisms for internal symmetries \cite{Horowitz:1989ng,MR1637718,Geiller:2020edh,Geiller:2020okp}.
In addition to the Lorentz transformations, we consider the so-called ``translations'' acting as
\be
\delta_\psi^\text{t}e=\de_\omega\psi,
\q\q
\delta_\psi^\text{t}\omega=\f{1}{\ell^2}[e,\psi],
\ee
where $\psi$ is a Lie algebra-valued 0-form. This is not an independent symmetry since on-shell the diffeomorphisms are given by field-dependent internal gauge transformations as $\pounds_\xi\approx\delta^\text{L}_{\xi\ip\omega}+\delta^\text{t}_{\xi\ip e}$. However, we are free to forget about the diffeomorphisms and study the asymptotic symmetries and charges of triad gravity in terms of the Lorentz transformations and translations. Since these symmetries generically depend on $3+3$ functions $\alpha$ and $\psi$ of all coordinates, it is a priori not clear that they carry the same physical content as the diffeomorphism charges \eqref{final charge}. We will show that this is however indeed the case, and in particular that the asymptotic internal symmetries preserving the triad are parametrized by four functions of $u$ and $\phi$.

We set out to determine the six gauge parameters $(\alpha^i,\psi^i)|_{i=0,1,2}$ such that $\delta_\epsilon=\delta^\text{L}_\alpha+\delta^\text{t}_\psi$ preserves the triad. In Bondi--Weyl gauge with the triad chosen to be \eqref{triad}, the six equations which determine the gauge parameters are solved in the following order:
\begin{enumerate}
\item The condition $(\delta_\epsilon e)_r^1=0$ fixes $\psi^1$ up to an $r$-independent integration function $\psi^1_0$.
\item The condition $(\delta_\epsilon e)_r^2=0$ fixes $\alpha^1$.
\item The condition $(\delta_\epsilon e)_\phi^1=0$ fixes $\psi^2$ up to an $r$-independent integration function $\psi^2_0$.
\item The condition $(\delta_\epsilon e)_u^1=0$ fixes $\alpha^2$.
\item The condition $(\delta_\epsilon e)_u^2=0$ fixes $\alpha^0$.
\item The condition that $(\delta_\epsilon e)_r^0$ be independent of $r$, which can be written $(\delta_\epsilon e)_r^0=b_0(u,\phi)$, fixes $\psi^0$ up to an $r$-independent integration function $\psi^0_0$.
\end{enumerate}
After fixing these conditions the transformation $\delta_\epsilon$ preserves $e$, and is parametrized by four arbitrary functions of $(u,\phi)$. In particular, we see that $b_0(u,\phi)$ is related to the field transformation $\delta_\epsilon\beta$.

This four-dimensional functional freedom at the end of the calculation reflects the four-dimensional freedom $(f,g,h,k)$ which parametrizes the asymptotic Killing vectors \eqref{AKV}. In fact the charges associated to the residual symmetries $\delta_\epsilon$ correspond to a change of slicing of the charges \eqref{final charge}. More precisely, after including the corner terms needed for renormalization and integrability, one can find the field-dependent redefinition of the free functions $(\psi^0_0,\psi^1_0,\psi^2_0,b_0)$ (these expressions are lengthy and we do not reproduce them here) such that the charges become exactly \eqref{Q total via EC} in the form
\be
\delta\Q=\oint_S(\alpha\delta e+\psi\delta\omega)+2\slashed{\delta}\big(\Q_\text{ren}+\Q_1\big)=\delta\big(\J_\alpha+\T_\psi+2\Q_\text{ren}+2\Q_1\big).
\ee
The resulting charge is \eqref{Q total via EC} where $\Q_\text{EC}+\J_\rho+\J_\lambda$ has been repackaged into $\J_\alpha+\T_\psi$. Notice that in the calculation of the asymptotic symmetries $\alpha$ and $\psi$ end up being field-dependent.

\section{Origin of the renormalization and integrability corner terms}
\label{sec:origin}

Let us now go back to the metric formulation. We have shown that the final expression \eqref{final charge} for the integrable and finite charge is\footnote{Our notation is chosen to remind the reader of the fact that none of the contributions on the right-hand side are integrable on their own, but that their sum is integrable, and gives of course $\delta\Q$.}
\be\label{final charge without c2}
\delta\Q=\slashed{\delta}\big(\Q_\text{EH}+\Q_\text{ren}+\Q_1).
\ee
The piece $\Q_\text{EH}$ has a covariant origin, since it comes from the contraction of the transformation $\pounds_\xi$ with the symplectic structure $\Omega_\text{EH}$. However the corners terms $\vartheta_\text{ren}$ and $\vartheta_1$ leading to $\Q_\text{ren}$ and $\Q_1$ are so far written only as variations of metric components (in \eqref{corners}) and do not have a covariant expression. In this section we ask the question of the covariant origin of the corner term $\vartheta_\text{ren}+\vartheta_1$. As surprising as it may seem, it turns out that this corner term is exactly the relative potential $\Omega_\text{EC/EH}$. From this observation we derive two equivalent results:
\begin{itemize}
\item[$i)$] One can work with $\Omega_\text{EC/EH}$ in its initial form, i.e. written in triad variables, but since the triad has an anomaly in the sense \eqref{triad anomaly}, we should use the field-space Lie derivative $\mathsf{L}_\xi$ to compute the charge.
\item[$ii)$] Alternatively, we can find a covariant rewriting of $\Omega_\text{EC/EH}$ in metric variables, which we will call $\Omega_\text{EC/EH}^\text{metric}$, at the price of introducing extra structure in the form of a vector. This then allows to use the spacetime Lie derivative $\pounds_\xi$ to compute the charge.
\end{itemize}
At the end of the day, we therefore show here that one can obtain \eqref{final charge without c2} by acting with the proper notion of Lie derivative on a covariant object. With either possibility listed above, we have
\bsub
\be
\delta\Q
&=-\mathsf{L}_\xi\ipp(\Omega_\text{EH}+\Omega_\text{EC/EH}),\\
&=-\pounds_\xi\ipp\big(\Omega_\text{EH}+\Omega_\text{EC/EH}^\text{metric}\big).
\ee
\esub
It is very important to notice here that we are \textit{adding} $\Omega_\text{EC/EH}$ (or its metric version) to $\Omega_\text{EH}$. This is therefore \textit{not} taking us back to the EC formulation (which would have required a subtraction). This should be clear from \eqref{Q total via EC}, which has already told us that even the EC formulation needs a corner term for renormalization and integrability.

\subsection{Corner with triad and anomaly}

To understand exactly in which sense the corner potential \eqref{DPS corner} ``knows'' about the renormalization and integrability corner terms, we can evaluate it on-shell. Noting that it is the component along the celestial circle which matters for our discussion, we find
\be\label{DPS on-shell}
(\vartheta_\text{EC/EH})_\phi\approx e^\varphi(r-H)\delta\beta=\vartheta_\text{ren}+\vartheta_1,
\ee
which is precisely the corner terms we have added to the metric formulation.\footnote{Note that $\vartheta_2$ and the associated charge $\Q_2$ will not play a role in this section. This is consistent because after all we have really added this term by hand for illustrative purposes in the previous sections, and we will see that there is indeed no relative corner term which controls its presence.} While this result is interesting because it tells us how to write the corner terms $\vartheta_\text{ren}+\vartheta_1$ in a covariant form using $\vartheta_\text{EC/EH}$, it is for the moment at odds with \eqref{QDPS}. It seems that computing the charges and evaluating the corner term on-shell does not commute! Indeed, the off-shell contraction \eqref{Lie in DPS} produces \eqref{QDPS}, while clearly if we first evaluate the corner potential on-shell and then use the right-hand side of \eqref{DPS on-shell} we obtain only $\slashed{\delta}\big(\Q_\text{ren}+\Q_1\big)$. The reason behind this apparent mismatch is precisely the lack of covariance of the triad in the sense \eqref{triad anomaly}. If, instead of the spacetime Lie derivative, we take into account $\Delta_\xi$ and use the field-space Lie derivative to compute the charges coming from $\vartheta_\text{EC/EH}$, we find consistently that
\be
-\mathsf{L}_\xi\ipp\Omega_\text{EC/EH}=\slashed{\delta}\big(\Q_\text{ren}+\Q_1\big).
\ee

Let us therefore go back to our question: Instead of adding the renormalization and integrability corner terms by hand, is there a covariant expression which can be added to the bulk potential such that the action of $\delta_\xi$ produces the total charge \eqref{final charge without c2}? The answer is affirmative provided we allow ourselves to use the triad while working in the metric formulation, and recall that its transformation is given by $\delta_\xi=\mathsf{L}_\xi$ and not $\pounds_\xi$ (while on the metric we have $\delta_\xi=\mathsf{L}_\xi=\pounds_\xi$ since it is covariant). The statement is then that\footnote{Once again, note that the symplectic structure in the bracket is \textit{not} $\Omega_\text{EC}$.}
\be\label{charge from covariant L}
\slashed{\delta}\Q=-\mathsf{L}_\xi\ipp(\Omega_\text{EH}+\Omega_\text{EC/EH}).
\ee
A natural question is whether this formula is merely a coincidence or if it holds for other gauges and solution spaces. That is to say, is the information about the (possible) renormalization and integrability corner terms always encoded in $\vartheta_\text{EC/EH}$, or is it a fluke of the Bondi--Weyl gauge? We keep this investigation for future work.

\subsection{Corner with metric and extra vector}

Another question, which we now turn to, is whether it is possible to write $\vartheta_\text{EC/EH}$ in terms of metric data instead of using the triad. Due to the form of \eqref{DPS corner}, which involves two components of $e$ and an inverse component $\hat{e}$, it is manifest that there is no direct metric expression. If a metric expression can be written at all, it must necessarily involve extra structure. We now show that this is the normal vector \eqref{v vector} introduced above.

To understand how this comes about, let us first recall that the Gibbons--Hawking--York (GHY) term $K=\nabla_\mu n^\mu$ can be written in triad variables using the internal normal $n^i=n^\mu e^i_\mu$ as
\be
\sqrt{-\gamma}\,K=([e,n]\wedge\de_\omega n)_{u\phi}=\big(\eps_{ijk}e^in^j(\de_\omega n^k)\big)_{u\phi}.
\ee
Now, following \cite{Wieland:2017zkf,Geiller:2017whh,Freidel:2020svx}, one can realize that this two-form, when seen as a boundary Lagrangian, contains itself a symplectic potential, i.e. a co-dimension two corner term. This corner term is
\be\label{EC/GR}
\vartheta_\text{EC/GR}=e[\delta n,n]=[e,\delta n]n=[n,e]\delta n,
\ee
and the associated symplectic current is $\delta\vartheta_\text{EC/GR}=\delta[n,e]\delta n=[\delta n,e]\delta n+[n,\delta e]\delta n$. From this we can compute that for Lorentz transformations we have
\be
-\delta_\alpha\ipp\Omega_\text{EC/GR}=-\J_\alpha=-\delta_\alpha\ipp\Omega_\text{EC/EH},
\ee
so that, just like the relative EC/EH corner term, the corner potential EC/GR removes the Lorentz charges coming from the triad formulation. The name of this corner term \eqref{EC/GR} comes form the fact that it is the relative potential between the triad Einstein--Cartan and the canonical metric ADM formulation of gravity. This latter was called GR in \cite{Freidel:2020xyx,Freidel:2020svx}. This will become manifest below.

To get a fully consistent picture, we can then relate $\vartheta_\text{EC/GR}$ and $\vartheta_\text{EC/EH}$ via a third relative potential, called $\vartheta_\text{GR/EH}$. This latter is the relative potential between the GR and EH formulations \cite{Harlow:2019yfa,Freidel:2020xyx}. It is the potential associated with the GHY boundary Lagrangian, and derived from the variational formula
\be\label{variation K}
\delta K=-\f{1}{2}K^{\mu\nu}\delta g_{\mu\nu}+\f{1}{2}\left(g^{\mu\nu}n^\alpha\nabla_\alpha-n^\mu\nabla^\nu\right)\delta g_{\mu\nu}-\f{1}{2}D_\mu(n^\alpha\gamma^{\mu\nu}\delta g_{\alpha\nu}).
\ee
In this expression, which is written in terms of the normal $n$ to slices at constant $r$, the corner in $D_\mu$ is a co-dimension two term which can be contracted with the null normal $s_\mu=(1,0,0)$ to obtain (after reintroducing the volume element and dropping a sign for convenience)
\be\label{HW corner}
\vartheta_\text{GR/EH}=\f{1}{2}\sqrt{-\gamma}\,s_\mu n^\alpha\gamma^{\mu\nu}\delta g_{\alpha\nu}=-\sqrt{-\gamma}\,s_\mu\delta n^\mu_\perp,
\q\q
\delta n^\mu_\perp\coloneqq\f{1}{2}(\delta n^\mu+g^{\mu\nu}\delta n_\nu),
\ee
which is (3.45) of \cite{Harlow:2019yfa} or (C.17) of \cite{Freidel:2020xyx} contracted with the normal $s_\mu$. 
%with the null normal $s_\mu=e^{2\beta}(1,0,0)\ \Rightarrow\ s^\mu=(0,-1,0)$ used in addition to $n^\mu$ to define
%\be
%(\vartheta_\text{GR/EH})_\mu=\f{1}{2}\sqrt{-\gamma}\,\eps_{\mu\nu\rho}s^\nu n^\alpha\gamma^{\rho\sigma}\delta g_{\alpha\sigma}=\f{1}{2}\sqrt{-\gamma}\,\eps_{\mu\nu\rho}s^\nu(2n^\rho\delta n^\sigma-\delta g^{\rho\sigma})n_\sigma,
%\ee
% This can also be written as
%\be
%(\vartheta_\text{GR/EH})_\mu=\f{1}{2}\sqrt{-g}\,\eps_{\mu\nu\rho}\tilde{s}^\nu n^\alpha\gamma^{\rho\sigma}\delta g_{\alpha\sigma}=\f{1}{2}\sqrt{-g}\,\eps_{\mu\nu\rho}\tilde{s}^\nu(2n^\rho\delta n^\sigma-\delta g^{\rho\sigma})n_\sigma,
%\ee
%with the normal $\tilde{s}$ defined as
%\be
%\tilde{s}_\mu=\f{\sqrt{-\gamma}}{\sqrt{-g}}e^{2\beta}(1,0,0),
%\q\q
%\tilde{s}^\mu\tilde{s}_\mu=0,
%\q\q\
%\tilde{s}^\mu n_\mu=-1.
%\ee
Consistently, one can finally use this to show the desired result, i.e. that we have\footnote{In these expressions the one-forms are understood as their $\phi$ component.}
\be\label{potentials chain rule}
\vartheta_\text{EC/EH}=\vartheta_\text{EC/GR}+\vartheta_\text{GR/EH}.
\ee
This is the chain-rule type of relation which connects all the relative corner potentials introduced so far.

The reason for which we have recalled these definitions and relations is that they will now enable us to write a metric equivalent of $(\vartheta_\text{EC/EH})_\phi$. First, inspired by \eqref{HW corner} we define the extra corner term
\be
\vartheta_\text{EC/GR}^\text{metric}\coloneqq\f{1}{2}\sqrt{-\gamma}\,s_\mu v^\alpha\gamma^{\mu\nu}\delta g_{\alpha\nu},
\ee
which is using $v$ instead of the normal $n$. To justify the name given to this corner term, we evaluate all the corner potentials on-shell to reveal that
\bsub\label{onshell corner potentials}
\be
(\vartheta_\text{EC/GR})_\phi&\approx\;2(\vartheta_\text{ren}+\vartheta_1)+\vartheta_2+T,\label{theta EC/GR}\\
\vartheta_\text{EC/GR}^\text{metric}&\approx\;2(\vartheta_\text{ren}+\vartheta_1)+\vartheta_2+T,\\
\vartheta_\text{GR/EH}&\approx-(\vartheta_\text{ren}+\vartheta_1)-\vartheta_2-T,
\ee
\esub
where $T$, whose explicit expression is not necessary, contains terms of order $\O(r^0)$ and subleading.
%\be
%T=\ell^2e^{\varphi-2\beta}\Big(\delta(U\varphi'+U')-2U'\delta\beta-2\delta\beta(U\varphi'+\partial_u\varphi)+\partial_u\delta\varphi\Big)-\delta(e^\varphi H)+\O(r^{-1}),
%\ee
This shows that $\vartheta_\text{EC/GR}^\text{metric}$ is indeed the metric equivalent of $(\vartheta_\text{EC/GR})_\phi$, and that, as announced, its construction has required the use of the extra structure $v$. Putting all this together, we can now generalize \eqref{DPS on-shell} to write
\be\label{DPS on-shell extended}
(\vartheta_\text{EC/EH})_\phi
&=(\vartheta_\text{EC/GR})_\phi+\vartheta_\text{GR/EH}\cr
&=\vartheta_\text{EC/GR}^\text{metric}+\vartheta_\text{GR/EH}\cr
&\approx e^\varphi(r-H)\delta\beta\cr
&=\vartheta_\text{ren}+\vartheta_1,
\ee
where one can notice that the corner potential $\vartheta_2$ appearing in \eqref{onshell corner potentials} has dropped. Instead of \eqref{DPS corner} we can therefore write a metric expression
\be
\vartheta_\text{EC/EH}^\text{metric}\coloneqq\f{1}{2}\sqrt{-\gamma}\,s_\mu(n^\alpha+v^\alpha)\gamma^{\mu\nu}\delta g_{\alpha\nu},
\ee
where it can be noted that
\be
n^\mu+v^\mu=\sqrt{-\f{\W}{r\B}}\,\partial_r.
\ee
As announced, we can then finally write \eqref{charge from covariant L} in the form
\be\label{charge from pounds}
\slashed{\delta}\Q=-\pounds_\xi\ipp\big(\Omega_\text{EH}+\Omega_\text{EC/EH}^\text{metric}\big).
\ee
The moot point is that there is a trade-off in ambiguities: either we use $\eqref{charge from covariant L}$ and introduce the operation $\mathsf{L}_\xi$, or we use \eqref{charge from pounds} and introduce the vector $v$.

\subsection{Boundary Lagrangians for the corner terms}

To wrap up, it would be interesting to follow the proposal of \cite{Harlow:2019yfa,Freidel:2020xyx,Freidel:2021cbc} and study whether the corner potentials actually descend from boundary Lagrangians. We have already explained that it is indeed the case for $\vartheta_\text{GR/EH}$ and $\vartheta_\text{EC/GR}$, which come respectively from the metric and triad GYH terms. We can then sum these two boundary Lagrangians so as to reproduce \eqref{potentials chain rule}. This means however that the boundary Lagrangian mixes different (metric and triad) variables. There is indeed no known pure triad boundary Lagrangian which gives rise to $\vartheta_\text{EC/EH}$. In light of the present discussion we believe that such an object does not exist, and that a correct boundary Lagrangian must necessarily use extra structure, such as \textit{both} a metric and a triad, or a metric and (in the present case) the vector $v$.

To write a co-dimension one Lagrangian whose symplectic potential is $\vartheta_\text{EC/EH}^\text{metric}$, we can first notice that a simpler on-shell expression is\footnote{We note that this ressembles the Hayward corner term \cite{Hawking:1996ww,Takayanagi:2019tvn,Freidel:2020xyx}, although here $\beta$ is \textit{not} a boost angle between normals.}
\be
\vartheta_\text{EC/EH}^\text{metric}\approx\sqrt{q}\,\delta\beta,
\ee
where $q\coloneqq\det(g_{\phi\phi})$, and $\beta$ is related to the volume of the normal metric $h_{ij}|_{i,j=u,r}$ in the null decomposition as $|\det(h_{ij})|=-g_{ur}=e^{2\beta}$. Since the $u$ component of the vector $v$ is given by $v^u=\sqrt{q}/\sqrt{-\gamma}$, our corner term can be obtained from the boundary Lagrangian $\sqrt{-\gamma}\,L_\beta=\sqrt{-\gamma}\,v^\mu D_\mu\beta$. Indeed, the variation of this latter produces a boundary term of the form $D_\mu(\sqrt{-\gamma}\,v^\mu\delta\beta)$, and on the co-dimension two corner the contraction with the normal $s$ complementing $n$ indeed gives
\be
s_\mu\sqrt{-\gamma}\,v^\mu\delta\beta=\sqrt{q}\,\delta\beta.
\ee
Alternatively, it is also interesting to notice that the boundary Lagrangian $L_0=-D_\mu v^\mu$ used in \eqref{boundary Lagrangian} contains a corner term which can be read from the variation
\be
\delta(\sqrt{-\gamma}\,L_0)=\delta\sqrt{-\gamma}\,L_0-\sqrt{-\gamma}\left(\delta\gamma^\mu_\alpha\nabla_\mu v^\alpha+D_\mu\delta v^\mu+\f{1}{2}\gamma^{\mu\sigma}v^\beta\nabla_\beta\delta g_{\sigma\mu}\right).
\ee
Contracting this corner term with $s_\mu$ gives, as one can check, the result
\be
-\sqrt{-\gamma}\,s_\mu\delta v^\mu=\vartheta_\text{EC/GR}^\text{metric}.
\ee
Putting this together tells us that the boundary Lagrangian $\sqrt{-\gamma}\,(L_\text{GHY}+L_0)$ gives rise to our corner potential following \eqref{DPS on-shell extended}.

Unfortunately, this argument is not airtight since neither $\sqrt{-\gamma}\,(L_\text{GHY}+L_0)$ nor $\sqrt{-\gamma}\,L_\beta$ are boundary Lagrangians which we have used to renormalize the on-shell action in section \ref{sec:renormalized action}. As mentioned above, this is due to the fact that we have not chosen a variational principle and therefore not uniquely fixed the form of the boundary Lagrangian. In this situation, it seems to us that it is not possible to fix the relationship between the corner terms and the boundary Lagrangian more than we have already done.

%\be
%\sqrt{q}\,\delta\beta=\f{1}{2}\sqrt{-\gamma}\,s_\mu(n^\alpha+v^\alpha)\gamma^{\mu\nu}\delta g_{\alpha\nu},
%\ee
%\be
%v^\mu\delta\beta=\f{1}{2}(n^\alpha+v^\alpha)\gamma^{\mu\nu}\delta g_{\alpha\nu},
%\ee
%Variation of the boundary Lagrangian
%\be
%\delta(\sqrt{-\gamma}L_b)
%&=\delta\sqrt{-\gamma}L_b+\ell\sqrt{-\gamma}(D_\sigma v^\sigma)\Big(\delta\gamma^\mu_\alpha\nabla_\mu v^\alpha+\gamma^\mu_\alpha\nabla_\mu\delta v^\alpha-\gamma^\mu_\alpha\delta\Gamma^\alpha_{\mu\beta} v^\beta\Big)\cr
%&=\delta\sqrt{-\gamma}L_b+\ell\sqrt{-\gamma}(D_\sigma v^\sigma)\left(\delta\gamma^\mu_\alpha\nabla_\mu v^\alpha+\gamma^\mu_\alpha\nabla_\mu\delta v^\alpha-\f{1}{2}\gamma^{\mu\sigma}(\nabla_\mu\delta g_{\sigma\beta}+\nabla_\beta\delta g_{\sigma\mu}-\nabla_\sigma\delta g_{\mu\beta})v^\beta\right)\cr
%&=\delta\sqrt{-\gamma}L_b+\ell\sqrt{-\gamma}(D_\sigma v^\sigma)\left(\delta\gamma^\mu_\alpha\nabla_\mu v^\alpha+\gamma^\mu_\alpha\nabla_\mu\delta v^\alpha-\f{1}{2}\gamma^{\mu\sigma}v^\beta\nabla_\beta\delta g_{\sigma\mu}\right)\cr
%&=\delta\sqrt{-\gamma}L_b+\ell\sqrt{-\gamma}(D_\sigma v^\sigma)\left(\delta\gamma^\mu_\alpha\nabla_\mu v^\alpha+D_\mu\delta v^\mu-\f{1}{2}\gamma^{\mu\sigma}v^\beta\nabla_\beta\delta g_{\sigma\mu}\right)\cr
%\ee

\section{Conclusion}
\label{sec:conclusion}

In this paper we have introduced a new gauge and solution space for three-dimensional gravity, motivated by the possibility of constructing non-trivial charges associated with Weyl rescalings of the boundary metric. We have done this construction in Bondi coordinates, so it is valid for any value of the cosmological constant and in particular in the flat limit $\ell\to\infty$. The construction builds up on previous work in three dimensions \cite{ruzziconi2020conservation} where the boundary metric is completely free and the variational principle is not fixed a priori. These are so-called leaky boundary conditions. In addition to this relaxation, we have considered a generalization of the so-called relaxed determinant condition \eqref{relaxed det} \cite{Barnich:2010eb}, and have considered instead \eqref{new relaxed det}. This introduces a new subleading function $H(u,\phi)$ whose role is crucial in the appearance of the Weyl charges.

On top of the usual ingredients of the covariant phase space formalism \cite{Lee:1990nz,Iyer:1994ys,Wald:1999wa,Barnich:2001jy,Barnich:2007bf}, the construction of finite and integrable charges has required to introduce corner terms for renormalization and integrability, and also to find an integrable slicing \cite{Adami:2020ugu,Adami:2021sko}. At the end of the day, in this slicing the charges are given by \eqref{final charge} (where as we have explained one can take $c_2=0$). Surprisingly, this contains four towers of $u$-dependent charges. The algebra is given by $\mathfrak{vir}\oplus\mathfrak{vir}\oplus\text{Heisenberg}$ and therefore contains three central extensions.

We have explained in details how the construction of the charges can be carried out in the triad formulation. This makes use of the relative corner term discussed in \cite{DePaoli:2018erh,Oliveri:2019gvm,Freidel:2020xyx}, but requires a careful analysis in order to understand that the renormalization and integrability corner terms are not the same in the metric and triad formulations. The discussion of the triad formulation has then enables us to explain in section \ref{sec:origin} the covariant origin of the renormalization and integrability corner terms.

These results suggest many possible directions for future work:
\begin{itemize}
\item To our knowledge, it is the first time that a relaxed determinant condition of the form \eqref{new relaxed det} is used. It would be very interesting to study this relaxation and its implications in higher-dimensional gravity.
\item In addition to the study of the metric and triad formulations we which carried out here, it would be interesting to reproduce these results in the Chern--Simons formulation. The reason for this is that eventually one would like to compare the gauge choices, solution spaces and the renormalization procedure which have been proposed in the literature with the construction of \cite{Grumiller:2016pqb,Grumiller:2017sjh}, which obtains in the Chern--Simons formulation six constrained towers of charges.
\item More generally, building up on the previous point, it would be interesting to understand what is the maximal number of finite and integrable charges which can be turned on in three-dimensional gravity (and of course also in higher-dimensional cases). It is noteworthy for example that an algebra similar to the $\big(\text{Diff}(C_2)\loplus\C^\infty(C_2)\big)\oplus\C^\infty(C_2)$ found here (which becomes $\mathfrak{vir}\oplus\mathfrak{vir}\oplus\text{Heisenberg}$ when represented in terms of the charges) has been found in \cite{Ciambelli:2021vnn}, however in a very different context and without a realization in terms of charges. Moreover, the algebra which we have found here turns out to be an extension by Weyl translations of the algebra derived around finite null hypersurfaces whose radial position is kept fixed \cite{Adami:2020ugu}. This could support the idea that the algebra which we have obtained here is maximal in the sense of \cite{Ciambelli:2021vnn} and of \cite{Adami:2020ugu} when allowing for Weyl translations, but then begs the question of the status of the six charges found in \cite{Grumiller:2016pqb,Grumiller:2017sjh}.
\item There is a known explicit diffeomorphism between the Bondi and the Fefferman--Graham gauges \cite{Poole:2018koa,Compere:2019bua,Ruzziconi:2020cjt,Ciambelli:2020ftk,Ciambelli:2020eba}. One interesting question is therefore how it can act on the Bondi--Weyl gauge and on the associated charges. Along the same line of thought one can wonder how the Bondi--Weyl gauge relates to other gauges, used for example in the study of holographic fluids \cite{Ciambelli:2020ftk,Ciambelli:2020eba}. 
\item In \cite{Adami:2020ugu}, boundary conditions around a finite null surface were considered. In the integrable slicing the algebra was found to be Diff$(S^1)\oplus$Heisenberg, where the Heisenberg part is however between the supertranslations and the Weyl dilatations. It would be interesting to study how this algebra found at finite null surfaces in \cite{Adami:2020ugu} is related to the algebra found here, and also if four towers of charges can be found at finite null surfaces.
\item It would be interesting to study further the holographic aspects of the Bondi--Weyl gauge, and in particular the boundary stress tensor, the Weyl anomaly, and possibly the boundary dynamics along the lines of \cite{Barnich:2012rz,Barnich:2013yka,Alessio:2020ioh}. In particular this could help understand the role of the family \eqref{boundary Lagrangian} of boundary Lagrangians which we have used to renormalize the action.
\item Finally, another possible generalization is to consider as the starting point the so-called Mielke--Baekler Lagrangian for three-dimensional gravity in triad variables \cite{Blagojevic:2004hj,Blagojevic:2005pd,Geiller:2020edh}. This Lagrangian contains a Chern--Simons term and a torsion term, which allows to obtain two different central charges for the Virasoro algebras (or the two central charges of BMS$_3$ in the flat limit). This would require to revisit the holographic renormalization of the potential, and also to add further corner terms for integrability.
\end{itemize}

\section*{Acknowledgement}
We would like to thank Romain Ruzziconi and Shahin Sheikh-Jabbari for discussions and comments. CZ also thanks Hamed Adami, Daniel Grumiller, Vahid Taghiloo and Hossein Yavartanoo for collaborations on related subjects.
CZ was supported by the Austrian Science Fund (FWF), projects P 30822 and M 2665. CG was supported by the Alexander von Humboldt Foundation.

\appendix

\section{Variations of $\boldsymbol{M}$ and $\boldsymbol{N}$}
\label{app:variations}

For the interested reader, we give here the form of the variations of the functions $M$ and $N$. In order to keep the expressions short, we given them in terms of components of the Killing equation at order $\O(r^0)$ and in terms of the other field variations \eqref{variations}. We have
\bsub
\be
\delta_\xi N&=e^{-2\beta}\Big((\pounds_\xi g)_{u\phi}\big|_{r^0}+\delta_\xi(e^{2\varphi}UH^2)\Big)-4(\delta_\xi H+2H\delta_\xi\beta)\beta'-2N\delta_\xi\beta,\\
\delta_\xi M
&=\f{1}{2}e^{-2\beta}(\pounds_\xi g)_{uu}\big|_{r^0}+\delta_\xi(UN)+4e^{2(\beta-\varphi)}(\beta')^2\delta_\xi(\varphi-2\beta)-\f{1}{2}e^{-2\beta}\delta_\xi(e^{2\varphi}U^2H^2)\cr
&\phantom{=\ }+2(UN+4UH\beta'-M)\delta_\xi\beta+4\beta'\delta_\xi(UH).
\ee
\esub

\addcontentsline{toc}{section}{References}

\providecommand{\href}[2]{#2}\begingroup\raggedright\endgroup

\end{document}